\documentclass[12pt]{article}
\usepackage{fullpage,amsmath, amsfonts, amssymb, amsthm, bm, natbib}
\usepackage[toc,page]{appendix}
\usepackage{bbold} % to write indicator function by \mathbb{1}
\usepackage{graphicx, algorithm, enumerate, enumitem, color, hyperref}
\usepackage[noend]{algpseudocode}  % to write pseudocode
\usepackage{bbm}
\usepackage{comment}
\newtheorem{remark}{Remark}
\usepackage{hyperref}

\newcommand{\real}{\mathbb R}
\newcommand{\keywords}[1]{\textbf{\small{Keywords:}} #1}

\def\Cov{\mathrm{Cov}}
\def\E{\mathrm{E}}
\def\Var{\mathrm{Var}}
\def\Prob{\mathrm{P}}
\def\P{\mathbb{P}}

\def\bnu{\bm{\nu}}
\def\bSigma{\bm{\Sigma}}
\def\bPsi{\bm{\Psi}}
\def\bPhi{\bm{\Phi}}
\def\bbeta{\bm{\beta}}
\def\bgamma{\bm{\gamma}}
\def\bdelta{\bm{\delta}}

\def\X{\bm{X}}

\def\Z{\bm{Z}}
\def\W{\bm{W}}

\def\A{\mathcal{A}}
\def\B{\mathcal{B}}

\def\I{\mathcal{I}}
\def\PP{\mathcal{P}}
\def\QQ{\mathcal{Q}}
\def\EE{\mathcal{E}}

\def\0{\bm{0}}
\def\1{\bm{1}}
\def\indi{\mathbb{1}}

\newtheorem{thm}{Theorem}
\newtheorem{lem}{Lemma}

\newtheorem{condition}{Condition}

\newcounter{example}[section]
\newcounter{question}[section]
\newcommand{\vertiii}[1]
{{\left\vert\kern-0.25ex\left\vert\kern-0.25ex\left\vert #1
\right\vert\kern-0.25ex\right\vert\kern-0.25ex\right\vert}}

\newcommand{\norm}[1]{\left\lVert#1\right\rVert}
\newcommand{\snorm}[1]{\|#1\|}

\DeclareMathOperator*{\argmin}{arg\,min}

\newlist{condenum}{enumerate}{1} % 'condenum': a new, enumerate-like list env.
\setlist[condenum]{label=\bfseries Condition \arabic*., 
                   ref=\arabic*, wide}

\graphicspath{ {./figures/} }

\title{\bf Reluctant Interaction Modeling in Generalized Linear Models}
\author{% begin a tabular environment to center authors as a block
\begin{tabular}{c}
Hai Lu \thanks{hailu@pstat.ucsb.edu} \qquad Guo Yu \thanks{guoyu@ucsb.edu} \\
Department of Statistics and Applied Probability \\
University of California Santa Barbara
\end{tabular}% end of the tabular environment
}
\date{}
\begin{document}
\maketitle

\begin{abstract}
While including pairwise interactions in a regression model can better approximate response surface, fitting such an interaction model is a well-known difficult problem. 
In particular, analyzing contemporary high-dimensional datasets often leads to extremely large-scale interaction modeling problem, where the challenge is posed to identify important interactions among millions or even billions of candidate interactions. While several methods have recently been proposed to tackle this challenge, they are mostly designed by (1) assuming the hierarchy assumption among the important interactions and (or) (2) focusing on the case in linear models with interactions and (sub)Gaussian errors. In practice, however, neither of these two building blocks has to hold. 
In this paper, we propose an interaction modeling framework in generalized linear models (GLMs) which is free of any assumptions on hierarchy. 
We develop a non-trivial extension of the reluctance interaction selection principle to the GLMs setting, where a main effect is preferred over an interaction if all else is equal. Our proposed method is easy to implement, and is highly scalable to large-scale datasets. 
Theoretically, we demonstrate that it possesses screening consistency under high-dimensional setting.
Numerical studies on simulated datasets and a real dataset show that the proposed method does not sacrifice statistical performance in the presence of significant computational gain.
\end{abstract}

\keywords{\small{high-dimensional interaction models, generalized linear models, variable screening}}

\section{Introduction}

Fitting a large regression model with interactions is a well-known computationally challenging task, as the number of candidate pairwise interaction grows quadratically with the number of main effects. Contemporary datasets are featured with a large number of predictors. 
Building a predictive model by including the pairwise interactions among these predictors then leads to a challenging variable selection problem with potentially more than millions of candidate interactions. 

Numerous methods have been proposed in the literature to tackle this challenging problem, including (among others):
\begin{itemize}
  \item Penalized regression based approaches: \citet{zhao2009composite, yuan2009structured, choi2010variable, radchenko2010variable, schmidt2010convex, bien2013lasso, lim2015learning, haris2016convex, she2018group,  hazimeh2020learning, tang2020high, zhang2023general};
  \item Multi-stage modeling approaches: \citet{wu2009genome,wu2010screen,hao2014interaction,shah2016modelling, hao2018model,2019arXiv190708414Y};
  \item Variable screening based approaches: \citep{li2014fast, fan2016interaction, niu2018interaction,reese2018strong, zhou2019bolt, xiong2023efficient, anzarmou2023kendall}.
  
\end{itemize}
However, virtually all of these methods are built on at least one of the following two crucial assumptions:
\begin{enumerate}
  \item From a structural point of view, a hierarchical assumption \citep{nelder1977reformulation, peixoto1987hierarchical} has been largely used, which requires that an interaction enters the model only if either or both corresponding main effects are already included in the model. While the hierarchical assumption facilitates computation by significantly narrowing down the search space of candidate interactions, the validity of this assumption has been questioned in many applications. For example, in human genetics, many genotypes have been detected to influence phenotype jointly rather than individually \citep{culverhouse2002perspective, cordell2009detecting, wan2010boost, fang2017tsgsis}. In this case, gene-gene interactions should be included in the model despite of the absence of their main effects.

  \item From a distributional point of view, it has been primarily assumed a (sub)Gaussian linear model with interactions for the responses, even if the hierarchical structure is not assumed \citep[see, e.g.,][]{fan2016interaction, niu2018interaction, 2019arXiv190708414Y}. Methods based on this assumption, as well as their theoretical properties, e.g., screening consistency \citep{radchenko2010variable, hao2014interaction}, hinge on the specific linear dependence of the response on main effects and interactions. As a result, these methods are not easily extendable to more general assumptions on the response distributions (e.g., binomial, Poisson, multinomial, ordinal, etc.).
\end{enumerate}

In this paper, we propose a computationally efficient method in interaction modeling that is both free of the hierarchy assumption and is applicable to a wide range of generalized linear models (GLMs). Such an interaction modeling method has rarely been studied in the literature. Our method is built on the \textit{reluctance principle} to interaction modeling, proposed in \cite{2019arXiv190708414Y} for (sub)Gaussian linear models with interactions, which claims that one should prefer main effects over interactions if all else is equal. However, the proposed method in \citet{2019arXiv190708414Y} hinges on the notion of residuals in linear models with interactions and cannot be easily extended to other GLMs. 

We develop a new strategy that honors the reluctance principle and can be applied to any GLMs. Specifically, we select interactions based on their contributions to the likelihood conditional on any arbitrary model fit using only the main effects. Our contributions and the organizations of the rest of the paper is as below:
\begin{enumerate}
  \item We introduce our reluctant interaction modeling framework in Section \ref{sec:proposal}, which is free of any structural assumptions (e.g., hierarchy) among interactions and applies to any GLMs. Specifically we define the set of \textit{pure interactions} that we aim at recovering, and establish that our framework includes \cite{2019arXiv190708414Y} as an example in linear models with interactions.
  \item In Section \ref{sec:sprinter} we formally introduce the algorithm that honors the proposed reluctant principle in GLMs, and highlight its computational efficiency and scalability to large-scale interaction modeling problems. For example, in Section \ref{sec:Time} we show that a model with around 2 million candidate interactions can be fitted in less than 30 seconds with a 5-fold cross-validation (on an Apple M1 CPU with 8GB RAM).
  \item In Section \ref{sec:theory} we show that the proposed method can simultaneously achieve computational efficiency and consistent recovery of the important interactions. These theoretical properties are further verified in simulation studies (Section \ref{sec:simu}) and a real dataset application (Section \ref{sec:Tripadvisor}).
\end{enumerate}

%Throughout the paper, we use $\X = (X_1, \ldots, X_p) \in \real^p$ to denote a generic $p$-dimensional feature vector. Let $q=(p^2 + p)/2$ be the total number of pairwise interactions among $p$ main effects, and $\Z = (X_1*X_1, X_1 * X_2, \ldots, X_p * X_p) \in \real^{q}$ denotes the vector of all pairwise interactions.
%We further let $[p] = \left\{ 1, 2, 3\dots, p \right\}$.
%Finally, we let $\tau: [p]^2 \rightarrow [q]$ map the interaction between $X_j$ and $X_k$ to its corresponding index in $\Z$, i.e., $Z_{\tau(j, k)} = X_j \ast X_k$.
%For simplicity, we denote $\gamma_{jk} = \gamma_{\tau(j, k)}$.
%({\color{cyan} how many of these notation is needed at all? How many of them is needed here?})

\section{Our proposal}\label{sec:proposal}
We consider a generic GLM, where the conditional distribution of the response $Y$ given main effects $\X \in \real^p$ belongs to an exponential family with the following density function
\begin{align}
  f(Y | \X, \theta) = \exp \left\{ Y \theta (\X) - b(\theta(\X)) + c(\X; Y) \right\},
  \label{mod:glm}
\end{align}
where $b(\cdot)$ and $c(\cdot)$ are known. Without loss of generality, throughout this paper, we suppose $\E(X_j) = 0$ and $\Var(X_j) = 1$, and we will focus on modeling pairwise interactions $\Z = (X_1*X_1, X_1 * X_2, \ldots, X_p * X_p) \in \real^q$ by 
\begin{align}
  \E [Y | \X] = b'(\theta(\X)) =  g^{-1} \left( \X^T \bbeta^\ast + \Z^T \bgamma^\ast \right),
  \label{mod:inter}
\end{align} 
where $\bbeta^\ast \in \real^p$ and $\bgamma^\ast \in \real^q$ are unknown but fixed coefficients for the main effects $\X$ and interactions $\Z$, respectively, $g(\cdot)$ is a known link function, and $q$ is the total number of interactions. 
For simplicity of presentation, we consider the canonical link function such that $g^{-1} = b'$, and thus model the mean as
\begin{align}
  \theta(\X) = \X^T \bbeta^\ast + \Z^T \bgamma^\ast.
  \label{mod:theta}
\end{align}
We note that while we focus on modeling the pairwise interactions, our method can be naturally extended to deal with higher order interactions.

\subsection{Reluctance to interactions in GLMs}\label{sec:reluctant}
One might naturally consider obtaining $\ell_1$-penalized maximum likelihood estimation of $\bbeta^\ast$ and $\bgamma^\ast$ simultaneously, a method to which we refer to as \textit{All pairs lasso} (\texttt{APL}).
However, \texttt{APL} could be computationally inefficient or even infeasible when the value of $p$ is large, as the number of coefficients grows quadratically with $p$. 
This computational limitation of \texttt{APL} stems from the fact that \texttt{APL} treats main effects and interactions equally. Statistically, by doing so, the successful selection of important main effects could be swamped by the overabundance of candidate interactions. 
%Finally, interactions are harder to interpret than main effects.

When an interaction has the same predictive power as a main effect, it could be beneficial to prefer the main effect over the interaction. While this desired reluctance to fitting an interaction is not utilized in \texttt{APL}, it is the main premise of the reluctant interaction modeling framework proposed in \cite{2019arXiv190708414Y}. 
Specifically, this proposed framework first captures the response signal using a model that is additive in the main effects, and then selects interactions that are highly correlated with the residual from the initial additive fit.
The marginal correlation between the residual and an interaction serves as a proxy for the \textit{purity} of that interaction, i.e., the additional predictive power to the response conditional on the best additive fit using only the main effects. 
By prioritizing fitting main effects over interactions, this approach is highly efficient in computation, and provides interpretable selection of interactions without requiring any hierarchical assumptions.
%Intuitively, interactions selected in this way are interpreted as interactions that cannot be captured by any linear combinations of main effects.

However, this framework hinges on the notion of residual and correlation, which do not easily extend to other GLMs. While there are indeed several analogs of residuals in GLMs \citep{pierce1986residuals}, in this paper we take a completely different approach that is based on likelihood.
Specifically, for each interaction $Z_j$, we consider the following maximum likelihood estimator 
\begin{align}
  (\check{\bbeta}^T, \check{\gamma}_{j})^T \in \argmin_{\bbeta \in \real^{p}, \gamma_{j} \in \real} \E \ell\left( \X^T \bbeta + Z_j \gamma_{j}, Y \right),
  \label{eq:marginal}
\end{align}
where the negative log-likelihood function in \eqref{mod:glm} is $\ell(\theta, Y) = b(\theta) - \theta Y$. 
Intuitively, $|\check{\gamma}_j|$ measures the contribution of the interaction $Z_j$ to predicting $Y$, conditional on a linear combination of main effects.
Selecting interactions based on $|\check{\gamma}_j|$ from \eqref{eq:marginal} thus exactly honors the reluctance principle, and falls into the framework of the conditional sure independence screening \citep[CSIS,][]{barut2016conditional}, where the conditional set is taken to be all main effects. 

There is, unfortunately, a clear computational shortcoming of this approach: \eqref{eq:marginal} requires solving a $(p + 1)$-dimensional optimization problem for each candidate interaction. Thus obtaining $|\check{\gamma}_j|$ for all $j \in [q]$ is computationally infeasible for large value of $p$.
%As an alternative approach, we apply reluctance principle on CSIS to propose a new interaction screening procedure called \textit{reluctant interaction screening}. As stated before, reluctance principle favors main effects over interactions for response signal capture. Therefore, we separate main effects and interactions fitting into two steps rather than fitting them together. 
Instead, we consider the following two-step approach, where we first use the main effects to capture the response signal as much as possible via
\begin{align}
  \bbeta^M \in \argmin_{\bbeta \in \real^p} \E \ell \left( \X^T \bbeta, Y \right),
  \label{eq:baseline}
\end{align}
and then add each $Z_j$ to the model for $j \in [q]$ conditional on the main effects fit:
\begin{align}
  \gamma^M_{j} \in \argmin_{\gamma_{j} \in \real} \E \ell \left( \X^T \bbeta^M + Z_j \gamma_{j}, Y \right).
  \label{eq:population}
\end{align}
The value $|\gamma^M_{j}|$ measures the marginal interaction signal that is not captured by the originally optimal main effects fit $\X^T \bbeta^M$. 
The following condition is required for the uniqueness of these quantities:
\begin{condition} \label{cond:unique}
  The population quantities \eqref{eq:marginal}, \eqref{eq:baseline}, and \eqref{eq:population} are unique.
\end{condition}

Notably, computing \eqref{eq:population} only involves an 1-dimensional optimization problem for each candidate interaction, and thus is highly scalable to large problems. 
The next section will demonstrate that this computational saving comes essentially free of concerns about any loss in statistical accuracy. Specifically, we will show that $\bgamma^M$ virtually encodes the same sparsity pattern as $\check{\bgamma}$, and thus can be used as a computationally efficient alternative to $\check{\bgamma}$ for the purpose of interaction selection.

\subsection{Pure interactions and their screening property}\label{sec:property}
\begin{thm} \label{thm:equivalence}
  Under Condition \ref{cond:unique}, for any $j \in [q]$, the following three statements are equivalent:
  \begin{enumerate}
    \item[S1.] $\check{\gamma}_j = 0$;
    \item[S2.] $\gamma^M_j = 0$;
    \item[S3.] $\Cov_L (Y, Z_j | \X) := \E \left[ \left( Z_j - \E_L (Z_j | \X) \right) \left( Y - \E_L(Y | \X) \right) \right] = 0$,
  \end{enumerate}
  where $\E_L (\cdot | \X)$ denotes the expectation linearly conditional on the main effects $\X$.
\end{thm}
\begin{proof}
See Appendix \ref{app:equivalence}.
\end{proof}

The equivalence between statement \texttt{S1} and statement \texttt{S2} in Theorem \ref{thm:equivalence} establishes the applicability of using the vector $\bgamma^M$ for screening out interactions: an irrelevant interaction in the sense of having $\check{\gamma}_j = 0$ is equivalently irrelevant in the sense of having $\gamma^M_j = 0$, while obtaining $\gamma_j^M$ only requires computing an $1$-dimensional (as opposed to a $(p + 1)$-dimensional) MLE.

The statement \texttt{S3} further characterizes the notion of an \textit{irrelevant interaction} (in the sense that $\check{\gamma}_j = 0$, and equivalently, $\gamma^M_j = 0$): any interaction $Z_j$ that is linearly uncorrelated with the response $Y$, conditional (linearly) on the main effects $\X$.
Here for a response variable $Y$, we define the linear (on main effects) conditional expectation $E_L(Y | \X)= b'(\X^T \bbeta^M)$, and for any interaction variable $Z_j$, we let $\E_L(Z_j | \X) = \Cov(Z_j, \X)\Cov(\X)^{-1} \X$.

Motivated Theorem \ref{thm:equivalence}, especially the equivalence to statement \texttt{S3}, in this paper we aim at recovering the set of \textit{pure interactions}, which we define as
\begin{align}
  \A(\kappa) = \left\{ j \in [q]: \left|\Cov_L \left( Y, Z_j | \X \right) \right| \geq n^{-\kappa} \right\},
  \label{def:set}
\end{align}
where the parameter $\kappa > 0$ characterizes the strength of the covariance (in absolute value) between the response and any pure interactions, conditional on the linear fit of main effects. This definition is highly interpretable, and perfectly aligns with the reluctant interaction principle: we only focus on interactions that provides additional predictive power to the response in the presence of the optimal linear fit of main effects. 

While virtually all of the interaction selection methods in literature focus on recovering the support set of the true interaction coefficients, i.e., $\mathrm{supp}(\bgamma^\ast) = \{j \in [q]: \gamma^\ast_j \neq 0\}$, we instead focus on recovering the set of pure interactions $\A(\kappa)$, which could be different from $\mathrm{supp}(\bgamma^\ast)$. Notably, in our framework, if the predictive power of $Z_j$ is well captured by the main effects (so that $\left|\Cov_L \left( Y, Z_j | \X \right) \right|$ is small), then it is of little interest to recover $Z_j$ even if $\gamma^\ast_j$ is significantly different from zero.

Despite the theoretical convenience in understanding the set of pure interactions, the definition in \eqref{def:set} provides little practical guidance as of how to recover such a set in a computationally feasible way. Just as in $\check{\gamma}_j$, obtaining $\Cov_L (Y, Z_j | \X)$ for each $j \in [q]$ effectively requires computing a $(p + 1)$-dimensional MLE. The following theorem bridges any pure interaction $j \in \A(\kappa)$ with its (nonzero) values of $|\gamma^M_j|$:

\begin{condition}\label{condition:b_prime}
  Let $m_j$ be the random variable defined by
  \begin{align}
    m_j = \frac{b' \left( \X^T \bbeta^M + Z_j \gamma^M_j \right) - b' \left( \X^T \bbeta^M \right)}{Z_j \gamma^M_j}.
    \label{eq:mj}
  \end{align}
  Then $c \E[m_j Z_j^2] \leq 1$ uniformly in $j \in [q]$ for some constant $c>0$.
\end{condition}

\begin{remark}
The strong convexity of $b(\cdot)$ implies that $m_j > 0$ almost surely. Moreover, if $b'(\cdot)$ is $\tau$-Lipschitz continuous (which we will assume later in Condition \ref{condition:tausmooth}), then $0 < m_j < \tau$, and thus Condition \ref{condition:b_prime} holds as long as $E[Z_j^2]$ is bounded, which holds if $\E[X_k^4]$ is bounded for $k \in [p]$ by Cauchy-Schwarz inequality. 
%To see this, note that
%\begin{align}
%  \E[Z_{\tau(j,k)}^2] = \E[X_j^2 X_k^2] \leq \sqrt{\E[X_j^4] \E[X_k^4]}.
%  \nonumber
%\end{align}
\end{remark}

\begin{thm} \label{thm:important}
  Under Condition \ref{condition:b_prime}, for the constant $c$ in Condition \ref{condition:b_prime}, we have
 \begin{align}
    \min_{j \in \A(\kappa)}|\gamma^M_j| \geq c \min_{j \in \A(\kappa)} \left| \Cov_L \left( Y, Z_j | \X \right) \right| \geq c n^{-\kappa}.
    \nonumber
  \end{align}
\end{thm}
\begin{proof}
See Appendix \ref{app:important}.
%  Denote
%  \begin{align}
%    \begin{pmatrix}
%      \Delta_{M, M} & \Delta_{M, j} \\
%      \Delta_{j, M} & \Delta_{j, j}
%    \end{pmatrix}:= 
%    \begin{bmatrix}
%      \E  m_j \X \X^T  &\E   m_j \X Z_j  \\
%      \E  m_j \X^T Z_j & \E  m_j Z_j^2. 
%    \end{bmatrix} 
%    \label{def:delta}
%  \end{align}
%  Using \eqref{eq:score_population} and \eqref{eq:score_baseline}, we have
%  \begin{align}
%    \E b' \left( \X^T \bbeta^M \right) \X = \E b' \left( \X^T \check{\bbeta} + Z_j \check{\gamma}_j \right) \X,
%    \nonumber
%  \end{align}
%  which implies that
%  \begin{align}
%    \E \left[ m_j \left( \X^T \check{\bbeta} + Z_j \check{\gamma}_j -  \X^T \bbeta^M \right) \right] \X = \0,
%    \nonumber
%  \end{align}
%  using the definition of $m_j$. This leads to
%  \begin{align}
%    \check{\bbeta} - \bbeta^M = - \Delta_{M, M}^{-1} \Delta_{M, j} \check{\gamma}_j
%    \label{eq:betadiff}
%  \end{align}
\end{proof}

Theorem \ref{thm:important} implies that the set of pure interactions $\A(\kappa)$ can be effectively defined (up to the constant $c$) in terms of the values of $|\gamma^M_j|$, which are much easier to compute than $\left| \Cov_L \left( Y, Z_j | \X \right) \right|$. Therefore, one can use the value $|\gamma^M_j|$, which is obtained by solving a $1$-dimensional MLE problem, as a powerful handle to recover $\A(\kappa)$. We propose our interaction selection algorithm based on $|\gamma^M_j|$ in Section \ref{sec:sprinter}.

\subsection{An application to the linear models}\label{sec:linear_app}
%\cite{fan2010sure} notes that the marginal maximum likelihood estimator (MMLE) of each variable is equivalent to its marginal correlation in linear models with normal errors.
In this section, we digress from the discussion of the proposed interaction modeling framework in any generic GLM, and focus specifically on a Gaussian linear model with pairwise interactions
\begin{align}
  Y = \X^T \bbeta^\ast + \Z^T \bgamma^\ast + \varepsilon, \quad \varepsilon \sim N(0, \sigma^2),
  \label{eq:normal}
\end{align}
which is simply an example of \eqref{mod:inter} with the identity link function under the Gaussian assumption. 
Our goal is to verify Theorem \ref{thm:equivalence} and Theorem \ref{thm:important} by evaluating the involved population quantities. The technical details are shown in Appendix \ref{app:linear_app_proof}.

We denote the three covariance matrices $\bSigma = \Cov(\X) \in \real^{p \times p}$, $\bPsi = \Cov(\Z) \in \real^{q \times q}$, and $\bPhi = \Cov(\X, \Z) \in \real^{p \times q}$. 
In \eqref{eq:normal}, the negative log-likelihood function $\ell(\theta, Y) \propto (Y - \theta)^2$, which leads to
\begin{align}
  \bbeta^M \in \argmin_{\bbeta \in \real^p} \E \left[ (Y - \X^T \bbeta)^2 \right]  = \bbeta^\ast + \bSigma^{-1} \bPhi \bgamma^\ast,
  \label{eq:lin_baseline}
\end{align}
and thus
\begin{align}
  \gamma^M_j \in \argmin_{\gamma \in \real} \E \left[ \left( Y - \X^T \bbeta^M - Z_j \gamma \right)^2 \right] = \Psi_{jj}^{-1} \left[ \bPsi_{j \cdot} - \bPhi_{j \cdot} \bSigma^{-1} \bPhi \right] \bgamma^\ast. 
  \label{eq:lin_population}
\end{align}
Importantly, it holds that
\begin{align}
  \check{\gamma}_j = \left( 1 - \Psi_{jj}^{-1} \bPhi_{j \cdot} \bSigma^{-1} \bPhi_{\cdot j} \right)^{-1} \gamma^M_j.
  \label{eq:check_bar}
\end{align}
In model \eqref{eq:normal}, Condition \ref{cond:unique} is equivalent to $\Psi_{jj} - \bPhi_{j\cdot} \bSigma^{-1} \bPhi_{\cdot j} \neq 0$, under which the equivalence of statement \texttt{S1} and statement \texttt{S2} in Theorem \ref{thm:equivalence} is verified.
Furthermore, it holds that
\begin{align}
  \Cov \left( Y, Z_j | \X \right) = \left[ \bPsi_{j \cdot} - \bPhi_{j \cdot} \bSigma^{-1} \bPhi \right] \bgamma^\ast = \Psi_{jj} \gamma^M_j, \nonumber
\end{align}
which verifies the equivalence of statement \texttt{S2} and statement \texttt{S3} in Theorem \ref{thm:equivalence} and defines the set of pure interactions as 
$$
\A(\kappa) = \left\{ j \in [q]:\Psi_{jj} \left| \gamma^M_j \right| \geq n^{-\kappa} \right\}.
$$
The definition in \eqref{eq:mj} further implies that $m_j = 1$. Consequently Condition \ref{condition:b_prime} posits $c \Psi_{jj} \leq 1$ for $j \in [q]$, and Theorem \ref{thm:important} is verified.

Finally we establish the equivalence to \citet{2019arXiv190708414Y}, which uses the following correlation (up to a constant) to select important interactions that cannot be captured by any main effects, i.e.,
\begin{align}
  \Psi_{jj}^{-1} \Cov(Z_j, Y - \X^T \bbeta^M) = \Psi_{jj}^{-1} \left[ \bPsi_{j \cdot} - \bPhi_{j \cdot} \bSigma^{-1} \bPhi \right] \bgamma^\ast = \gamma^M_j.
  \nonumber
\end{align}
In other words, in Gaussian linear models with interactions \eqref{eq:normal}, the marginal correlation ranking framework in \citet{2019arXiv190708414Y} is a special example of our proposed selection framework based on the value of $|\gamma^M_j|$.

%{\color{magenta} 
%$$
%\mathcal{I}(\alpha) \in \underset{\mathcal{A} \subseteq[q]}{\arg \max }\left\{\min _{\ell \in \mathcal{A}}\left|\Psi_{\ell \ell}^{-1 / 2} \operatorname{Cov}\left(Z_{\ell}, W^T \gamma^*\right)\right| \quad \text { s.t. } \quad\left\|\left(W_{\mathcal{A}^C}^T \gamma_{\mathcal{A}^C}^*\right)^2\right\|_{\psi_{1 / 2}} \leq \alpha\right\} .
%$$
%$$
%\mathcal{I}(\alpha) \in \underset{\mathcal{A} \subseteq[q]}{\arg \max }\left\{\min _{\ell \in \mathcal{A}}\left|\gamma^M_j\right| \quad \text { s.t. } \quad\left\|\left(W_{\mathcal{A}^C}^T \gamma_{\mathcal{A}^C}^*\right)^2\right\|_{\psi_{1 / 2}} \leq \alpha\right\} .
%$$
%$$\alpha^{1/2} = 2 \tilde{c} n^{4-\kappa}\|x\|_{\psi_2}^2$$
%}

\section{The \texttt{sprinter} algorithm in GLMs}\label{sec:sprinter}
In this section, we introduce Algorithm \ref{alg:sprinter}, named \texttt{sprinter} (for \textbf{sp}arse \textbf{r}elucant \textbf{inter}action) in GLMs, that honors the reluctant principle to interactions in Section \ref{sec:reluctant}. 
Consider a generic random vector $(U, V)$, and $n$ independent samples $\{(U_i, V_i)\}_{i = 1}^n$ that follow the same distribution as $(U, V)$. 
For any function $f$, we denote $\P_n f(U, V) = n^{-1} \sum_{i = 1}^n f(U_i, V_i)$.
%$\E[f(U, V)]$ to be the expectation of $f(U, V)$ taken w.r.t the joint distribution of $(U,V)$ (assuming the existence of the expectation). 

{\centering
  \begin{minipage}{.99\linewidth}
    \begin{algorithm}[H]
      \caption{sprinter}
      \begin{algorithmic}[100]
        \Require a log-likelihood function $\ell(\cdot, \cdot)$, penalty functions $\PP(\cdot)$ and $\QQ(\cdot)$, and a tuning parameter $\eta > 0$.
        \State \textbf{Step 1}: Following \eqref{eq:baseline}, obtain 
        \begin{align}
          \hat{\bbeta} \in \argmin_{\bbeta \in \real^p} \left[ \P_n \left\{ \ell \left( \X^T \bbeta, Y \right) \right\} + \PP(\bbeta) \right].
          \label{eq:step1}
        \end{align}
        \State \textbf{Step 2}: For each candidate interaction $Z_j$, following \eqref{eq:population} we compute 
        \begin{align}
          \hat{\gamma}_j = \argmin_{\gamma_j \in \real} \left[ \P_n \left\{ \ell \left(\X^T \hat{\bbeta} + Z_j \gamma_j, Y \right) \right\}  \right].
          \label{eq:gammahat}
        \end{align}
        \State \textbf{Step 3}: Select interactions based on $|\hat{\gamma}_j|$ and $\eta$:
        \begin{align}
          \hat \I_\eta = \left\{j \in [q]: \left|\hat{\gamma}_j \right| > \eta \right\}.
          \label{eq:Ihat}
        \end{align}
        \Return $\hat{\I}_\eta$
       \State \textbf{Step 4 (optional)}: Refit based on the previous main effects fit $\hat{\bbeta}$, using the main effects and the selected interactions 
        \begin{align}
          (\hat{\boldsymbol{\theta}}, \hat{\boldsymbol{\delta}}) \in \argmin_{\boldsymbol{\theta} \in \real^p, \boldsymbol{\delta} \in \real^{|\hat{\I}_\eta|}} \left[ \P_n \left\{ \ell \left(\X^T \hat{\bbeta} + \X^T \boldsymbol{\theta} + \Z_{\hat{\I}_\eta}\boldsymbol{\delta}, Y \right) \right\} + \QQ(\boldsymbol{\theta}, \boldsymbol{\delta}) \right] 
          \label{eq:refit}
        \end{align}
        \Return $(\hat{\bbeta} + \hat{\boldsymbol{\theta}},  \hat{\boldsymbol{\delta}})$
      \end{algorithmic}
      \label{alg:sprinter}
    \end{algorithm}
  \end{minipage}
  
}
\vspace*{0.3cm}
In Algorithm \ref{alg:sprinter}, obtaining $\hat{\bbeta}$ in Step 1 and $\hat{\gamma}_j$ in Step 2 are the empirical equivalents of computing the population quantities $\bbeta^M$ in \eqref{eq:baseline} and $\gamma_j^M$ in \eqref{eq:population}, respectively. Specifically, for high-dimensional main effects, i.e., $p$ is large, one could use a penalized MLE (with a user specified penalty $\PP$) for $\bbeta^M$ in Step 1. While for the simplicity of presentation we focus on using only the main effects in Step 1, in practice one could effectively use any $O(p)$ predictors that are derived from main effects, e.g., basis expansions or trees.

The utility of each interaction, $|\hat{\gamma}_j|$, is then computed as an 1-dimensional MLE in Step 2, and is then used in Step 3 to select pure interactions defined in \eqref{def:set} by Theorem \ref{thm:important}. The value of $\eta$ is a tuning parameter in this algorithm. Changing the value of $\eta$ is equivalent to varying the size of selected interactions set. In practice, for any given positive integer value $m$, we consider the following equivalence of \eqref{eq:Ihat}:
\begin{align}
\hat{\I}_m^{\text{top}} =\left\{j \in[q]: \left|\hat{\gamma}_j \right| \text { is among the top } m\text { largest}\right\}.
\label{eq:topm}
\end{align}
This \textit{top-$m$} approach is widely used in variable screening literature \citep[see, e.g.,][]{fan2008sure, fan2010sure, fan2016interaction, niu2018interaction, 2019arXiv190708414Y}, where the value of $m$ are commonly set to be $n$ or $[n/\log(n)]$ instead of tuning using data-adaptive approaches. 
%The theoretical screening property of $\hat{\I}_k^{\text{top}}$ is shown in Appendix \ref{app:topk}. 

If the goal is to further build a predictive model with the selected interactions, then one could take the optional Step 4 that jointly refits all the main effects and the selected interactions in $\hat{\I}_\eta$ in addition to the main effects fit $\X \hat{\bbeta}$ in Step 1. Since $p$ and (or) $|\hat{\I}_\eta|$ could be large, a user specified penalty $\QQ(\cdot, \cdot)$ could be used to obtain a regularized estimate as in \eqref{eq:refit}.

Our theoretical analysis is essentially free of specific penalty functions used in Step 1 and Step 3. In the numeric studies in this paper, we use $\ell_1$-penalty for linear, logistic, Poisson, multinomial, and ordinal logistic regression \citep[ordinalNet,][]{wurm2017regularized}. 
As illustrated in Section \ref{sec:linear_app}, on the population level, our interaction selection framework reduces to \cite{2019arXiv190708414Y} in Gaussian linear models with interactions. We note that Algorithm \ref{alg:sprinter}, with $\ell(\theta, Y) = (\theta - Y)^2$, is indeed Algorithm 1 in \citet{2019arXiv190708414Y}.

The worst-case computational complexity of Algorithm \ref{alg:sprinter} largely depends on Step 2 and Step 3, which requires a single pass over $O(p^2)$ candidate interactions to select top $m$ elements. Specifically, for each interaction,
computing \eqref{eq:gammahat} typically uses the Newton-Raphson algorithm, which take $O(n\log \epsilon)$ time to get an $\epsilon$-accurate solution, which leads to the $O(p^2 n\log \epsilon)$ time and $O(n)$ storage complexity.
Using the top-$m$ strategy in \eqref{eq:topm} in Step 3, we construct and maintain a min-heap to update the $m$ interactions to keep in $\hat{\I}_m^{\text{top}}$ as the search continues, which takes $O(p^2\log m)$ computation and $O(m)$ storage.
For a selected set of interactions of size $m$ obtained in Step 3, the dimensionality of the interaction space is effectively reduced from $O(p^2)$ to $O(m)$, which significantly lessens the computational burden for any downstream statistical tasks.
%If we apply lasso in Step 1 and Step 3, the whole time complexity for \textit{sprinter} is $O(p^2(\log k + n)) + O(n(p+k)\min\{n, p+k\})$. The overall storage needed is only $O(n(p+k))$. 
While the worst-case computational complexity of Algorithm \ref{alg:sprinter} is the same as that of \texttt{APL} (both scale with $p^2$), we note that Algorithm \ref{alg:sprinter} only needs a single pass over all candidate interactions, compared with multiple passes required for \texttt{APL} and other optimization based interaction selection methods. As a numeric comparison study between \texttt{sprinter} and \texttt{APL} shown in Section \ref{sec:Time}, we notice a significant improvement in terms of computation efficiency while not losing any statistical performance.

\section{Theoretical analysis}\label{sec:theory}
In addition to the computational advantages, we study the theoretical properties of the proposed framework. Specifically, in Section \ref{sec:consistency}, we focus on the selection properties of \texttt{sprinter}, i.e., if the set of pure interactions in $\A(\kappa)$ is recovered by $\hat{\I}_\eta$ for certain value of $\eta$. Of course, for the sole purpose of having $\A(\kappa) \subseteq \hat{\I}_\eta$, one could naively use $\eta = 0$ and thus select all possible interactions, which has no practical guidance. In Section \ref{sec:reduction}, we further quantify the size of the selected interaction set $\hat{\I}_\eta$ to ensure computation efficacy for the specific value of $\eta$ that recovers $\A(\kappa)$. In other words, our goal in this Section is to theoretically illustrate that \texttt{sprinter} enjoys computational savings without compromising any statistical performance. The technical proof of all the theoretical results are detailed in appendices.

By Condition \ref{cond:unique}, the unique minimizer $\bbeta^M$ and $\gamma^M_j$ are interior points of some sufficiently large convex and compact sets. Specifically, we denote $\B = \{a: |a| \leq B\}$ for some sufficiently large constant $B > 0$, and require that $\beta^M_k \in \B$ for all $k \in [p]$ and $\gamma^M_j \in \B$ for all $j \in [q]$.

\subsection{Selection consistency}\label{sec:consistency}
The selection consistency of \texttt{sprinter} is established based on the following conditions:
%\begin{condition} \label{condition:fisher}
%  For each $j \in [q]$, consider the Fisher information $$I(\gamma_{j}) = \E\left(b^{\prime \prime}(\bm{X}^{T} \bm{\beta}^{M} + Z_{j}\gamma_{j}) Z_j^2\right),$$
%  then
%  $\sup_{\gamma_{j} \in \B, |Z_j|=1}I(\gamma_{j})$ is bounded.
%\end{condition}

\begin{condition} \label{condition:tail}
  For each main effect $X_j$ for $j \in [p]$, there exists some positive constants $r_{0}$, $r_{1}$, $s_{0}$, $s_{1}$, and $\alpha$ 
  such that 
  \begin{align}
    \Prob \left(\left|X_j \right|>t\right) \leq r_{1} \exp \left(-r_0 t^{\alpha}\right),
    \label{eq:tail}
  \end{align}
  and for sufficiently large $t$,
  \begin{align}
    & \E \exp \left(b\left(\X^{T} \bbeta^{\ast}+\Z^{T} \bgamma^{\ast}+s_{0}\right) -b\left(\X^{T} \bbeta^{\ast}+\Z^{T} \bgamma^{\ast}\right)\right) \nonumber\\
    +& \E \exp \left(b\left(\X^{T} \bbeta^{\ast}+\Z^{T} \bgamma^{\ast}-s_{0}\right)
    -b\left(\X^{T} \bbeta^{\ast}+\Z^{T} \bgamma^{\ast}\right)\right) \leq s_{1}, \nonumber
  \end{align}
  where $({\bbeta^\ast}^T, {\bgamma^\ast}^T)^T$ is the true parameter in \eqref{mod:theta}.
\end{condition}

\begin{condition} \label{condition:tausmooth} 
  The function $b(\cdot)$ is $\tau$-smooth. In other words, $b'$ is $\tau$-Lipschitz continuous, i.e.,
  \begin{align}
    \left| b'(x) - b'(y) \right| \leq \tau \left| x - y \right|
    \nonumber
  \end{align}
  holds for any $x$ and $y$.
\end{condition}

\begin{condition} \label{condition:lip}
  The second derivative of $b(\cdot)$ is continuous and positive. There exists an $\varepsilon > 0$ such that for all $j \in [q]$ and any $a_0 \in \B$,
  \begin{align}
    \sup_{\gamma_j \in \B, \, |\gamma_j - \gamma^M_j| \leq \varepsilon} \left| \E \left[ b(a_0 + Z_j \gamma_j) \indi \left(  |Z_j| > K_n \right)\right] \right| = o(n^{-1}),
  \end{align}
  for some large constant $K_n$. Furthermore, let
  \begin{align}
    \Lambda_j = \left\{ (z_j, y): |z_j| \leq K_n, |y| \leq K_n^\ast \right\},
    \label{eq:lambdaj}
  \end{align}
  where $K_n^\ast = r_0 K_n^\alpha / s_0$ with constants $r_0$, $s_0$, and $\alpha$ as in Condition \ref{condition:tail}. The negative log-likelihood function $\ell(a_0 + z_j \gamma_j, y)$ is Lipschitz continuous in $\gamma_j$ for any $\gamma \in \B$, i.e., for any $a_0, \gamma_j, \tilde{\gamma}_j \in \B$,
  \begin{align}
    \left|\ell(a_0 + z_j \gamma_j, y) - \ell(a_0 + z_j \tilde{\gamma}_j, y) \right| \indi \left\{ (z_j, y) \in \Lambda_j \right\}
    \leq k_n |z_j (\gamma_j - \tilde{\gamma}_j)|\indi \left\{ (z_j, y) \in \Lambda_j \right\},
  \label{lipsch}
  \end{align}
  where the Lipschitz constant is
  \begin{align}
    k_n = b'(B (K_n + 1)) + r_0 K_n^\alpha / s_0.
    \label{eq:kn}
  \end{align}
\end{condition}

\begin{condition} \label{condition:lowerbound} 
  For any $\gamma_j$ such that $\left|\gamma_{j}-\gamma_{j}^{M}\right| \leq b_n$, we have
  \begin{align}
    \E\left[ \ell \left(\X^T \bbeta^M + Z_j\gamma_j, Y\right) - \ell \left(\X^T \bbeta^M + Z_j\gamma^M_j, Y\right) \right] \geq V(\gamma_j-\gamma^M_j)^2
    \nonumber
  \end{align}
  for some positive $V$, bounded from below uniformly over $j\in [q]$.
\end{condition}
%Condition \ref{condition:lowerbound} states that the deviation of negative likelihood expectation on $\gamma_j$ from its minimal value obtained on $\gamma_j^M$ is controlled by the difference between $\gamma_j$ and $\gamma_j^M$. 

Condition \ref{condition:tail}, \ref{condition:lip} and \ref{condition:lowerbound} are standard in the literature of variable screening in GLMs \citep[see, e.g., ][]{fan2010sure, barut2016conditional, fan2016interaction}. Specifically, 
%Condition \ref{condition:fisher} ensures the {\color{cyan}identifiability and the existence} of $\gamma^M_j$ for each $j$ and is satisfied in many GLMs ({\color{cyan} like what? Can we have a section in appendix giving some examples?}). 
the first inequality in Condition \ref{condition:tail} quantifies the distribution of the main effects through their tail behaviors. For example, generic sub-Gaussian random variables have $\alpha=2$ and their pairwise interactions could be sub-Exponential random variables.  Our theoretical results have explicit dependence on the value of $\alpha$, and thus allow various distributions of the main effects (and thus their interactions).
%Furthermore, Condition \ref{condition:tail} imposes the upper bound of $|z_j|$ and $|y|$ in set $\Lambda_j$ for Condition \ref{condition:lip}.
The $\tau$-smooth in Condition \ref{condition:tausmooth} and positive $b''(\cdot)$ in Condition \ref{condition:lip} together imply the strong convexity of $b(\cdot)$ with $0 < b''(\cdot) \leq \tau$, which enables Condition \ref{condition:lowerbound} to be satisfied in many GLMs. Specifically, we have $b''(\cdot) = 1$ in linear models with Gaussian errors and $0 < b''(\cdot) \leq 1/4$ in logistic regressions. Moreover, Condition \ref{condition:lip} and \ref{condition:lowerbound} specify the constraints on the negative log-likelihood function $\ell(\theta, Y)$. In Appendix \ref{app:kn} we provide reasoning for the choice of Lipschitz constant in \eqref{eq:kn}.

We provide a roadmap for establishing the pure interaction selection consistency. Recall that our goal is to establish that $\A(\kappa) \subseteq \hat{\I}_\eta$ with high probability for some suitably chosen value of $\eta$.
In order for all pure interactions in $\A(\kappa)$ to be detected, the signal strength $n^{-\kappa}$ in the definition of $\A(\kappa)$ in \eqref{def:set} should be large enough; specifically larger than certain \textit{noise level}. As we will characterize later in this section, this noise level is essentially $\max_j |\hat{\gamma}_j - \gamma^M_j|$, i.e., the error of $\hat{\gamma}_j$ in \eqref{eq:gammahat} for estimating its population quantity $\gamma^M_j$ in \eqref{eq:population}. 

The multi-step nature of \texttt{sprinter} suggests that $\max_j |\hat{\gamma}_j - \gamma^M_j|$ depends on two sources of errors: the model misspecification error in Step 1 by totally ignoring the interaction terms, and the concentration performance of the sample quantity around its mean in Step 2. 
Specifically, as we allow for user-specified approach in fitting \eqref{eq:step1} in Step 1 of \texttt{sprinter}, we denote its predictive performance as
\begin{align}
  \E \left[\left(\X^T ( \hat{\bbeta} - \bbeta^M ) \right)^2 \right] = \Theta(n^{-2\rho})
  \label{eq:rho}
\end{align}
for some $\rho \leq 1/2$, and establish subsequent results explicitly dependent on $\rho$. 
We then extend the analysis in \cite{fan2010sure} and \citet{barut2016conditional} to provide the concentration error rate. By controlling these two sources of error, we have the following main result:
\newcommand{\thmref}[2]{%
  \ref{#1}(S\ref{#2})%
}
\begin{thm}[Selection consistency] \label{thm:screening}
  Suppose that Condition \ref{condition:tail} to Condition \ref{condition:lip} hold with $r_0$, $r_1$, $s_1$, $k_n$, and $K_n$, and assume that $\E[X_j^4] \leq 4$ for all $j \in [p]$. Then for the set of pure interaction $\A(\kappa)$ defined in \eqref{def:set} where $\kappa < \rho$, and with $K_n^{-2}\min\{k_n^{-2} n^{1 - 2\kappa}, n^{2 \rho - 2 \kappa}\}$ $\rightarrow \infty$ as $n \rightarrow \infty$, we have
  \begin{enumerate}
    \item[S1.] For any constant $c_1 > 0$, there exists a universal constant $c_2 > 0$ such that
      \begin{align}
        &\Prob \left[\max_{j \in [q]} |\hat{\gamma}_j - \gamma_j^M| \geq c_1 n^{-\kappa}\right] \nonumber \\
        \leq &q \exp \left( - \frac{c_2}{K_n^2} \min \left\{ k_n^{-2} n^{1 - 2 \kappa}, n^{2 \rho - 2 \kappa} \right\}\right) + n q \left[s_1 \exp (-r_0 K_n^\alpha) + 2 r_1 \exp (-r_0 K_n^{\alpha / 2})\right].
        \nonumber
      \end{align}

    \item[S2.]  If in addition, Condition \ref{condition:b_prime} holds with constant $c$. Take $\eta = c_3 n^{-\kappa}$ with $c_3 \leq c / 2$, we have
      \begin{align}
        \Prob \left[ \A(\kappa) \subseteq \hat{\I}_\eta \right] 
        \geq &1 - |\A(\kappa)| \exp \left( - \frac{c_2}{K_n^2} \min \left\{ k_n^{-2} n^{1 - 2 \kappa}, n^{2 \rho - 2 \kappa} \right\}\right) \nonumber \\ & -  n|\A(\kappa)| \left[s_1 \exp (-r_0 K_n^\alpha) + 2 r_1 \exp (-r_0 K_n^{\alpha / 2})\right].
        \nonumber
      \end{align}
      
  \end{enumerate}
\end{thm}
\begin{proof}
  See Appendix \ref{app:proof_screening}.
\end{proof}

The results in Theorem \ref{thm:screening} depend on the values of model parameters $\rho$, $\kappa$, and $\alpha$. 
The value of $\rho$ quantifies the model misspecification error incurred in Step 1, due to the fact that the interaction signal is totally ignored on purpose to honor the reluctant interaction principle.
The value of $\kappa$ is the signal strength of pure interactions defined in \eqref{def:set}. 
The constraint that $\kappa < \rho $ indicates that the strength of pure interactions that can be recovered hinges on the predictive performance of Step 1.
Finally, the value of $\alpha$ characterizes the tail behavior of the distribution of main effects in Condition \ref{condition:tail}. For example, with $\alpha = 2$ the main effects are sub-Gaussian and their pairwise interactions could thus be sub-Exponential random variables. Heavier-tailed main effects leads to slower convergence of the probability (to one) with which the screening consistency result (Statement S2) holds.

In both logistic regression and linear regression models with pairwise interactions, we derive the order of $q$, i.e., the number of interactions that can be recovered by the proposed method, as a function of $n$ by ensuring that the probability with which Statement S2 in Theorem \ref{thm:screening} converges to one as $n$ approaches infinity.
The details of derivations can be found in Appendix \ref{app:Kn}.

Specifically, for logistic regression models, with $K_n = O(n^\frac{\min\{1 - 2\kappa, 2\rho - 2 \kappa\}}{2+\alpha/2})$, the probability lower bound in Statement \texttt{S2} is maximized, which implies that the scale of the number of interactions that the proposed method can successfully recover is $ \log(q) = o(n^\frac{\min\{1 - 2\kappa, 2\rho - 2 \kappa\}\alpha}{4+\alpha})$. 

In linear models with pairwise interactions, we have that
for $\alpha \geq 1$,
\begin{align}
\log (q)= \begin{cases}
o(n^\frac{(1-2\kappa)\alpha}{4+5\alpha}) & \text{if }  \kappa \leq (\frac{1}{\alpha} + \frac{5}{4}) (\rho - \frac{1}{2}) + \frac{1}{2}\\
o(n^{\frac{(2\rho-2\kappa)\alpha}{4+\alpha}}) & \text{if } \kappa > (\frac{1}{\alpha} + \frac{5}{4}) (\rho - \frac{1}{2}) + \frac{1}{2}
\end{cases},
\nonumber
\end{align} 
and for $\alpha < 1$,
\begin{align}
\log (q)= \begin{cases}
o(n^\frac{(1-2\kappa)\alpha}{8+\alpha}) & \text{if } \kappa \leq (\frac{\alpha}{4}+2)(\rho - \frac{1}{2}) + \frac{1}{2}\\
o(n^{\frac{(2\rho-2\kappa)\alpha}{4+\alpha}}) & \text{if } \kappa > (\frac{\alpha}{4}+2)(\rho - \frac{1}{2}) + \frac{1}{2}\\
\end{cases}.
\nonumber
\end{align} 
Regardless of the value of $\alpha$, the scale of the ambient dimension (the total value of candidate interactions $q$) in which the proposed method can recover the set of pure interactions depends on the signal strength of such pure interactions: if one aims at recovering only strong pure interactions (defined in \eqref{def:set} with small value of $\kappa$) , then the proposed method can handle much larger number of candidate interactions. On the other hand, if the ambition is to recover weaker pure interactions, then the problem dimension that the proposed method can handle is correspondingly smaller. 

The dependence of the target pure interaction signal $\kappa$ depends on both the Step 1 prediction error (through the value of $\rho$) as well as the main effects tail behaviors (through the value of $\alpha$). In particular, a necessary condition for non-negative threshold values of $\kappa$ is that $\rho > 5/18$, thus requiring Step 1 prediction error rate not to be too slow.
In addition, we observe an interesting phase transition phenomenon based on the tail behaviors of the main effects through the value of $\alpha$ in Condition \ref{condition:tail}. Specifically, the dependence of $\kappa$ on $\alpha$ as well as the dimension $q$ that can be handled by the proposed method are different, depending on whether the main effects tail behavior is slower than that of sub-exponential ($\alpha = 1$) random variables.

Finally, in linear models with pairwise interactions and sub-Gaussian main effects (where $\alpha = 2$), the results above imply that the proposed method is able to recover strong pure interactions among $q = \exp(o(n^\frac{1 - 2 \kappa}{7}))$ candidate interactions. 
This result is weaker than that provided in \cite{2019arXiv190708414Y}, where $\log(q) = \min\{o(n^{\min\{\frac{3 - 4 \kappa}{5}, 1-2\kappa\}}), O(n^{\frac{1}{2}})\}$. This difference stems from the fact that our proposed framework accommodates various GLMs, whereas the techniques in \cite{2019arXiv190708414Y} are specifically designed for linear models with pairwise interactions. 
%We also note that our results are weaker than those in \cite{barut2016conditional}, which develops a general conditional sure independence screening framework and can recover the set \eqref{} dimension of order $o(n^\frac{1 - 2 \kappa}{4})$ with variables that follow Condition \ref{condition:tail}. This difference happens because we focus on interactions, which have heavier behavior than main effects.

\begin{comment}
Plug in $k_n$, we have
\begin{align}
    1 & = O(\frac{1}{K_n^{2+\alpha/2}} \min \left\{ (B (K_n + 1) + K_n^\alpha / s_0) ^{-2} n^{1 - 2 \kappa}, n^{2 \rho - 2 \kappa} \right\}) \nonumber \\ 
      & = O(\frac{1}{K_n^{2+\alpha/2}} \min \left\{ \frac{n^{1 - 2 \kappa}}{\max\{K_n^{2},K_n^{2\alpha}\}}, n^{2 \rho - 2 \kappa} \right\})  \nonumber \\
       & = O(\min \left\{ \frac{n^{1 - 2 \kappa}}{\max\{K_n^{4+\alpha/2},K_n^{2+5\alpha/2}\}}, \frac{n^{2 \rho - 2 \kappa}}{K_n^{2+\alpha/2}} \right\})  \nonumber \\ 
      & = O(\min \left\{ \frac{n^{1 - 2 \kappa}}{K_n^{\max\{4+\alpha/2,2+5\alpha/2\}}}, \frac{n^{2 \rho - 2 \kappa}}{K_n^{2+\alpha/2}} \right\}) \nonumber
 \end{align}
 Therefore, $K_n \in \{ O(n^\frac{{1 - 2 \kappa}}{\max\{4+\alpha/2,2+5\alpha/2\}}), O(n^\frac{{2 \rho - 2 \kappa}}{2+\alpha/2})\}$
\end{comment}

\subsection{Dimension reduction efficacy}\label{sec:reduction}
As discussed, the pure interaction selection consistency is less helpful unless the size of the selected interactions can be effectively controlled. In the following theorem, we characterize the size of the retained interactions under the following additional condition:
\begin{condition} \label{condition:size} 
  \begin{enumerate}[label=(\arabic*)]
    \item The variance of interaction signal $\Var(\Z^T \bgamma^\ast) = O(1)$.
    \item $\min_{j \in [q]}\E(m_jZ_j^2) \geq C$ for some constant $C>0$, where $m_j$ is defined in Condition \ref{condition:b_prime}.
    \item Define 
    \begin{align}
      \bnu = \E \left[ \E_L[\Z | \X] (\X^T \bbeta^\ast + \Z^T \bgamma^\ast - \X^T \bbeta^M) \right] ,
      \label{def:w}
    \end{align}
It holds that $\snorm{\bnu}_2^2 = o (\lambda_{\max} (\bSigma_{\Z | \X}))$, where $\lambda_{\max} (\bSigma_{\Z | \X})$ is the largest eigenvalue of $\bSigma_{\Z | \X} := \E\{ [\Z - E_{L}(\Z | \X)] [\Z - E_{L}(\Z | \X)]^T \} $.
  \end{enumerate}
\end{condition}

In part (3) of Condition \ref{condition:size}, $\bnu$ quantifies the relationship between $\E_L[\Z | \X]$, which encapsulates how much the interactions $\Z$ can be predicted linearly from the main effects $\X$, and $\X^T \bbeta^\ast + \Z^T \bgamma^\ast - \X^T \bbeta^M$, which represents the interaction signal that the main effects fail to capture. The size of $\bnu$ is assumed to be small relative to $\lambda_{\max }\left(\bSigma_{\Z | \X}\right)$, the greatest variance in the interactions $\Z$ that remain after the optimal linear combination of main effects $\X$ has been removed. Particularly in linear models with pairwise interactions, where the link function $b^{\prime}(\theta)=\theta$, it follows from Appendix \ref{app:equivalence} that $\E \left[ \X \X^T \bbeta^M   \right]  = \E \left[\X (\X^T \bbeta^\ast + \Z^T \bgamma^\ast)\right]$. In this case, $\bnu=0$, and therefore part (3) of Condition \ref{condition:size} is trivially satisfied.

\begin{comment}
$\lambda_{\max} (\bSigma_{\Z | \X})$ is the largest eigenvalue of $\bSigma_{\Z | \X}$.
This eigenvalue represents the largest variance in the direction of the principal component of the interaction effects after accounting for the main effects.
\end{comment}

\begin{thm}\label{thm:size}
  Under Condition \ref{condition:b_prime} to Condition \ref{condition:size}, we have for the same choice of $\eta$ as in Theorem \ref{thm:screening} that
  \begin{align}
    &\Prob \left[ | \hat{\I}_\eta | = O(n^{2\kappa} \lambda_{\max}(\bSigma_{\Z | \X})) \right]  \nonumber\\
    \geq &1- q \exp \left( - \frac{c_2}{K_n^2} \min \left\{ k_n^{-2} n^{1 - 2 \kappa}, n^{2 \rho - 2 \kappa} \right\}\right) + n q \left[s_1 \exp (-r_0 K_n^\alpha) + 2 r_1 \exp (-r_0 K_n^{\alpha / 2})\right].
    \nonumber
  \end{align}
\end{thm}
\begin{proof}
  See Appendix \ref{app:size}.
\end{proof}
It shows that the number of selected interaction set is on the scale $O(n^{2\kappa} \lambda_{\max}(\bSigma_{\Z | \X}))$, which depends on the conditional covariance $\bSigma_{\Z | \X}$. For example, if $\X\sim\mathcal{N}\left(\mathbf{0}_{p},\mathbf{I}_{p}\right)$, we have $\lambda_{\max}(\bSigma_{\Z | \X}) = \lambda_{\max}(\E(\Z\Z^T)) = 3$, and consequently $| \hat{\I}_\eta | = O(n^{2\kappa})$. In such an example, the constraint $\kappa < 1/2$ implies that $|\hat{\I}_\eta | = o(n)$, which is on a significantly smaller order of $O(p^2)$ for \texttt{APL}.

\section{Simulation studies}\label{sec:simu}
\subsection{Logistic regression}\label{sec:logi}
Through simulation studies, we study the performance of the proposed \texttt{sprinter} algorithm in several GLMs.
We start with logistic regression models, which employ the logit link in \eqref{mod:inter}:
\begin{align}
\operatorname{logit}[\operatorname{Pr}(Y=1|\X)]=\X^T \bbeta^\ast + \Z^T \bgamma^\ast
\label{mod:logit}
\end{align}

For the implementation of the \texttt{sprinter} algorithm (i.e., Algorithm \ref{alg:sprinter}), we apply an $\ell_1$-penalty in both Step 1 and Step 4, and select the top-$m$ interactions with $m=[n/\log(n)]$ in Step 3. Additionally, we use a 5-fold cross-validation procedure to simultaneously select the tuning parameter pair for the $\ell_1$-penalties in Steps 1 and 4. The \href{https://github.com/hluscience/sprintr}{implementation of the \texttt{sprinter} algorithm} and the \href{https://github.com/hluscience/reproducible/tree/main/sprintr}{code that reproduces all of the subsequent numerical studies} are available in the first author's \href{https://github.com/hluscience}{GitHub repositories}.
We further include the following methods for comparison:
\begin{itemize}
    \item The Main Effects Lasso (\texttt{MEL}) applies the $\ell_1$-penalty to the model that includes only main effects. It is implemented using \texttt{glmnet}, with the tuning parameter selected by cross-validation.
    \item The All Pairs Lasso (\texttt{APL}) applies the $\ell_1$-penalty to the model containing both main effects and all the pairwise interactions. It also uses \texttt{glmnet} with the tuning parameter selected by cross-validation.
    \item The \texttt{SIS} employs sure independence screening \citep{fan2010sure} on pairwise interactions only, then applies the $\ell_1$-penalty to the model with all main effects and selected interactions. The tuning parameter is chosen by cross-validation in \texttt{glmnet}.
    
    \end{itemize}
    In addition, we include the following methods that require the hierarchical assumptions:
\begin{itemize}   
    \item \texttt{RAMP} \citep{hao2018model}, which adaptively grows interactions on a regularization path under the (weak) hierarchy principle. This is implemented using the \texttt{RAMP} package.
    \item \texttt{glinternet} \citep{lim2015learning}, which learns interactions via hierarchical group-lasso regularization. This is implemented by \texttt{glinternet} package.
    \item \texttt{hierNet} \citep{bien2013lasso}, which captures interactions through convex constraints with (weak) hierarchy restriction. This is implemented via \texttt{hierNet} package.
    \end{itemize}

\begin{comment}
We generate $n=1000$ independent observations from $\mathcal{N}\left(\mathbf{0}_{p},\mathbf{I}_{p}\right)$ with $p=150$. We use 10\% of the observations for the training dataset and the remaining 90\% for the evaluation dataset.
\end{comment}

We generate $n=100$ independent observations from $\mathcal{N}\left(\mathbf{0}_{p},\mathbf{I}_{p}\right)$ with $p=150$. Denote $\mathcal{M}$ as the indices of the main effects coefficient $\beta^\ast$ with non-zero coefficients and $\mathcal{I}$ as the indices of true interactions coefficient $\gamma^\ast$ with non-zero coefficients, we consider the following three structures of the true interactions:
\begin{enumerate}
\item Mixed: $\mathcal{M}=\{1,2,3\}$ and
$\mathcal{I}=\{(1,4),(2,5),(6,7),(8,9),(10,11)\}$.
\item (Weak) Hierarchical, i.e., $\gamma_{j k}^\ast \neq 0 \Longrightarrow \beta_{j}^\ast \neq 0$ or $\beta^\ast_{k} \neq 0$: $\mathcal{M}=\{1,2,3\}$ and $\mathcal{I}=\{(1,3),(1,4)$, $(2,5), (3,6),(1,7)\}$.
\item Anti-hierarchical, i.e., $\gamma_{j k}^\ast \neq 0 \Longrightarrow \beta^\ast_{j}=0$, and $\beta^\ast_{k}=0$: $\mathcal{M}=\{1,2,3\}$ and $\mathcal{I}=\{(4,5),(6,7)$, $(8,9), (10,11), (12,13)\}$.
\end{enumerate}
For $j\in \mathcal{I}$ we fix $\gamma_j^\ast = 4$, and for $j \in \mathcal{M}$ we vary the common value of $\beta_j^\ast \in \{1, 1.5, 2, 2.5, 3\}$. By doing so, the underlying signal has different main effects to interaction signal ratios. As the reluctant interaction principle prioritizes main effects over interactions, it is expected that \texttt{sprinter} will outperform in the settings with large values of $\gamma^\ast_j$, i.e., high main effects to interaction signal ratios.

\begin{figure}[H]
\begin{center}
\includegraphics[width=1\linewidth]{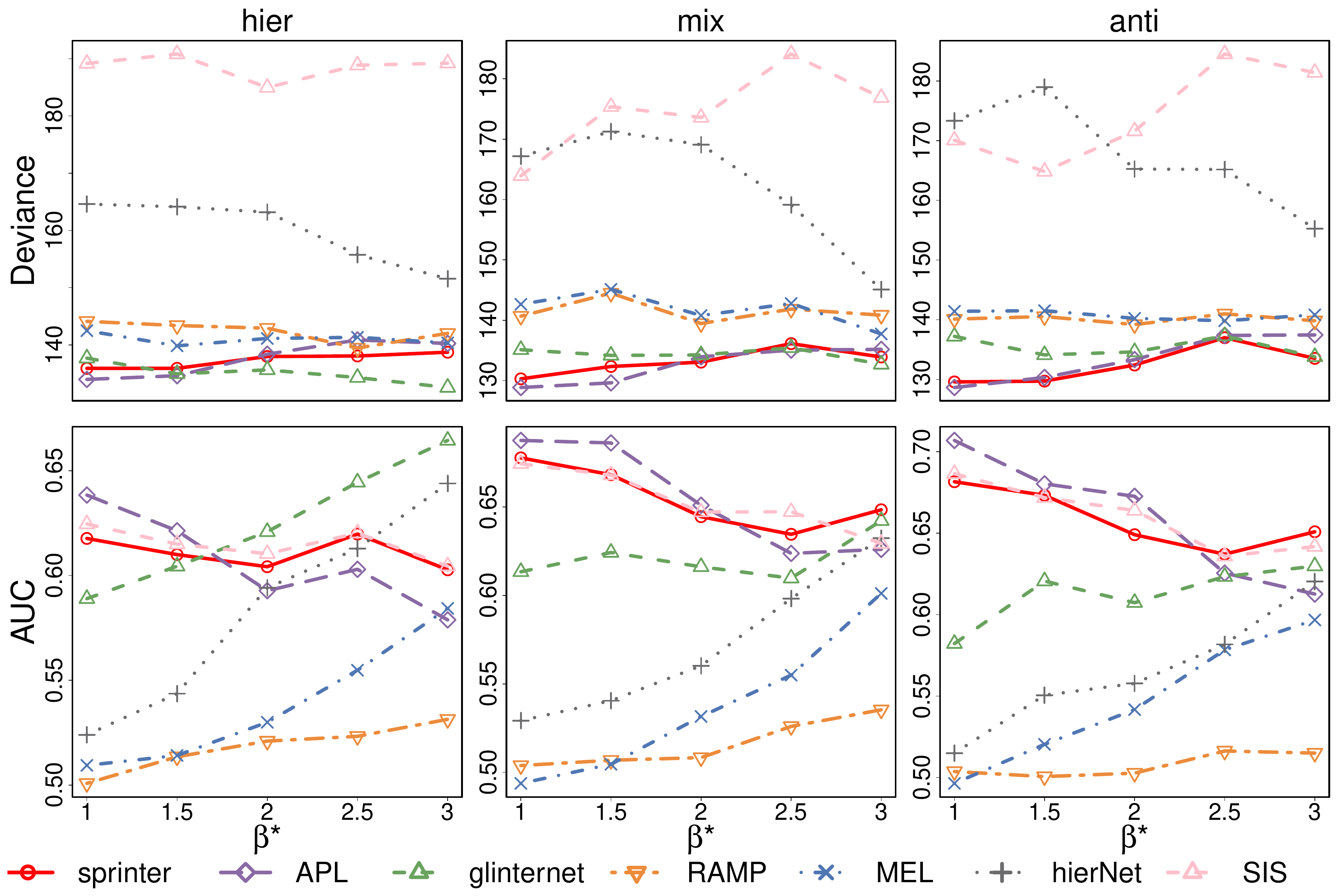}
\caption{Logistic regression models: deviance and AUC of various methods in comparison in the three interaction structure settings (averaged over 50 replications).}
\label{fig:simu}
\end{center}
\end{figure}

We evaluate the predictive performance of different methods using both deviance and AUC (from the ROC curves of the predicted response) on a separate independent evaluation dataset, which contains 100 observations that have the same distribution as the training data. Finally, all simulation studies are replicated $50$ times. The results are summarized in Figure \ref{fig:simu}.

\begin{comment}
\begin{figure}[H]
\centering
\includegraphics[width=.335\textwidth]{hier.logistic.dev.png}\hspace*{-1cm}\hfill
\includegraphics[width=.335\textwidth]{mix.logistic.dev.png}\hspace*{-1cm}\hfill
\includegraphics[width=.335\textwidth]{anti.logistic.dev.png}
\includegraphics[width=1\textwidth]{title.log.png}
\caption{Deviance of different methods in 3 different interaction structure scenarios (averaged over 50 simulations, logistic regression setting)}
\label{fig:simu}
\end{figure}
\end{comment}

We observe that \texttt{sprinter} performs consistently well, compared with other methods, in different simulation settings. Specifically, methods under hierarchical assumptions, i.e., \texttt{glinternet}, \texttt{hierNet}, and \texttt{RAMP}, work well when the underlying true interactions are indeed hierarchical, but their performance start deteriorating as the true interaction structure is deviating from hierarchy, and as the main effect coefficients $\beta^\ast$ are weak compared with the interaction coefficients $\gamma^\ast$.
In addition, \texttt{MEL} performs effectively as Step 1 in Algorithm \ref{alg:sprinter}. So the comparison between \texttt{MEL} and \texttt{sprinter} indicates the importance of effectively modeling the interactions. And as expected, such an improvement is mostly pronounced when $\beta^\ast$ is small. Just as \texttt{sprinter}, \texttt{APL} consistently performs well in three different settings of interaction structures. Compared with \texttt{sprinter}, \texttt{APL} performs strongly when the interaction signal dominates, but falls short particularly when $\beta^\ast$ gets larger. 
This reflects the fact that \texttt{APL} models the main effects and the interactions on the equal footing, while \texttt{sprinter} prioritizes main effects over interaction --- a strategy that is particularly effective when the true main effects dominate. 
We further demonstrate this key difference between \texttt{sprinter} and \texttt{APL} in Section \ref{sec:Time}, where we gradually increase the value of $p$, and observe that the performance of \texttt{APL} gets swamped by the tremendous number of candidate interactions. \texttt{SIS} performs well in terms of AUC, but significantly underperforms in terms of deviance. Our method consistently performs well in both criteria.

\subsection{Poisson regression}

\begin{comment}
in order to generate varying main-interaction-signal-ratios, which are measure by MIR
$$
    \operatorname{MIR}=\frac{\left\|\mathbf{X} \beta^*\right\|_2^2}{\left\|\mathbf{Z} \gamma^*\right\|_2^2}
$$
Similar to MAR, large MIR value indicates the scenarios that the signal of the main effects dominates the signal of the interaction effects, and oppositely, small MIR value characterizes the scenarios that the signal of the interaction effects dominate the signal of main effects. 

\begin{figure}[H]
\centering
\includegraphics[width=.335\textwidth]{hier.poi.png}\hspace*{-1cm}\hfill
\includegraphics[width=.335\textwidth]{mix.poi.png}\hspace*{-1cm}\hfill
\includegraphics[width=.335\textwidth]{anti.poi.png}
\caption{Comparison of deviance of various methods in 3 different interaction structure scenarios. In each interaction structure scenario, all the methods are compared on various MIR. x-axis is MIR, and y-axis is the deviance averaged over 100 replications on poisson regression setting under the MIR.}
\label{fig:poisson simu}
\end{figure}

Figure \ref{fig:poisson simu} shows the average deviance value from 100 simulations for each specified MIR value. We see that the proposed \textit{sprinter} is competitive with other methods under different structures with different main-interaction ratios.
\end{comment}
In poisson regression models, we use the logarithm link function in \eqref{mod:inter} and have 
\begin{align}
\operatorname{log} E(Y) =\X^T \bbeta^\ast + \Z^T \bgamma^\ast.
\label{mod:poisson}
\end{align}
While Possion regression is arguably the most commonly used GLMs for modeling responses that are count data, there is no implementations for \texttt{RAMP}, \texttt{glinternet}, and \texttt{hierNet} in their corresponding \texttt{R} packages. So we compare the performance of the proposed \texttt{sprinter} algorithm with \texttt{MEL}, \texttt{APL}, and \texttt{SIS}.

We generate $n=100$ independent observations from $\mathcal{N}\left(\mathbf{0}_{p}, 0.5 \mathbf{I}_{p}\right)$ with $p=150$. The small variance value of 0.5 is chosen to prevent unnecessarily large response values. As in Section \ref{sec:logi}, we consider the following three structures of the true interactions:

\begin{enumerate}
\item Mixed: $\mathcal{M}=\{1,2\}$ and $\mathcal{I}=\{(1,3), (4,5), (6,7)\}$.
\item (Weak) Hierarchical: $\mathcal{M}=\{1,2\}$ and $\mathcal{I} = \{(1,2), (1,3), (2,4)\}$.
\item Anti-hierarchical: $\mathcal{M}=\{1,2\}$ and $\mathcal{I}=\{(3,4), (5,6), (7,8)\}$.
\end{enumerate}
For $j\in \mathcal{I}$ we fix $\gamma_j^\ast = 2$, and for $j \in \mathcal{M}$ we vary the common value of $\beta_j^\ast \in \{1, 1.25, 1.5, 1.75, 2, 2.25\}$. 

\begin{figure}[H]
\begin{center}
\includegraphics[width=1\linewidth]{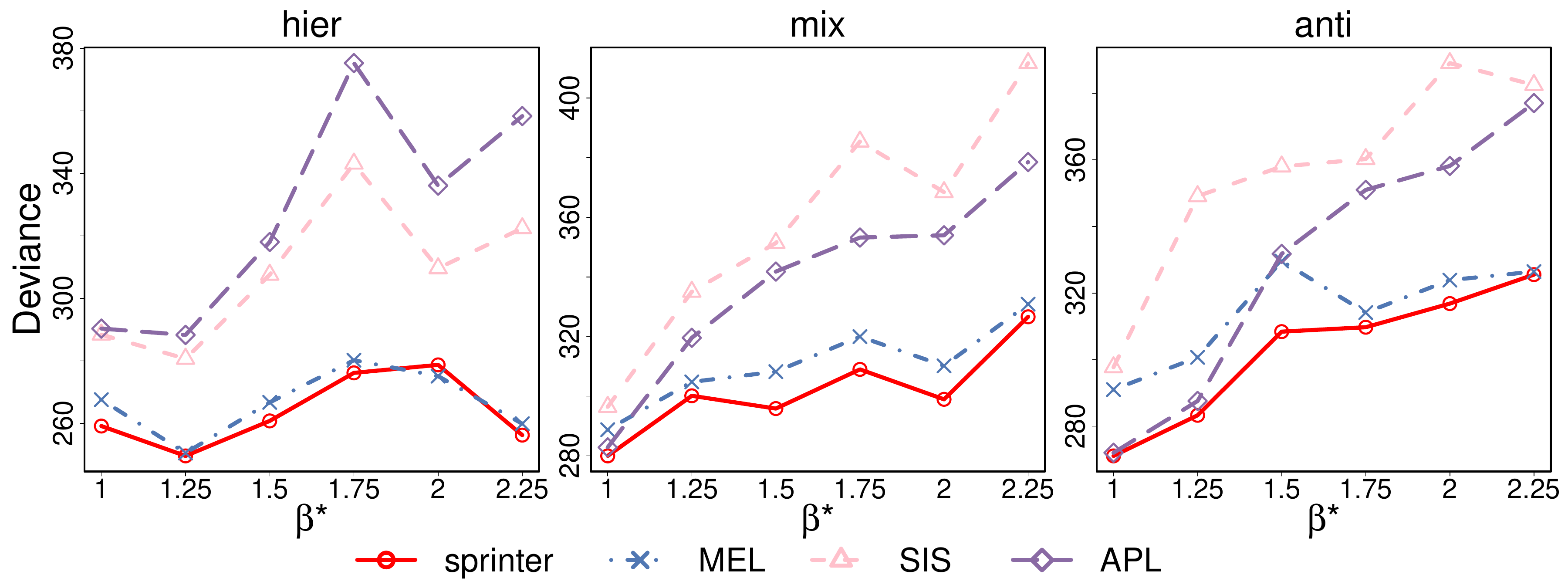}
\caption{Poisson regression models: deviance of various methods in comparison in the three interaction structure settings (averaged over 50 replications).}
\label{fig:poisson simu}
\end{center}
\end{figure}

We observe again that \texttt{sprinter} performs consistently better in comparison with other methods. Interestingly, the favorable performance of \texttt{MEL}, compared with that of \texttt{SIS} and \texttt{APL}, suggests that these are challenging settings where the error propagated through the process of selecting interactions (as in \texttt{SIS} or \texttt{APL}) overwhelms the model-misspecification error of totally ignoring interactions (as in \texttt{MEL}). 
Compared with \texttt{MEL}, \texttt{sprinter} shows the benefit of correctly selecting interactions in prediction.

\subsection{Comparison with \texttt{APL}}\label{sec:Time}
In this section, we focus on the logistic regression setting, and compare \texttt{sprinter} with \texttt{APL} in terms of both computational time and predictive performance (measured in deviance and AUC). Specifically, we generate $n=100$ observations as in Section \ref{sec:logi} with mixed structure with $\beta_j^* = 4$ and $\gamma_j^\ast = 4$, and we vary $p\in \{150,500,1000,1500, 2000\}$.

For these much larger problem sizes, we adapt a slightly different strategy for selecting tuning parameters in Algorithm \ref{alg:sprinter} than in the previous simulation studies: we first select and fix the tuning parameter for the $\ell_1$-penalty in Step 1 before proceeding to the subsequent steps. In particular, in Step 3, we retain the top-$m$ interactions where $m=[n/\log(n)]$ and in Step 4, we use a 5-fold cross-validation procedure to select the tuning parameter for the $\ell_1$-penalty.

\begin{figure}[H]
\begin{center}
  \includegraphics[width=1.03\linewidth]{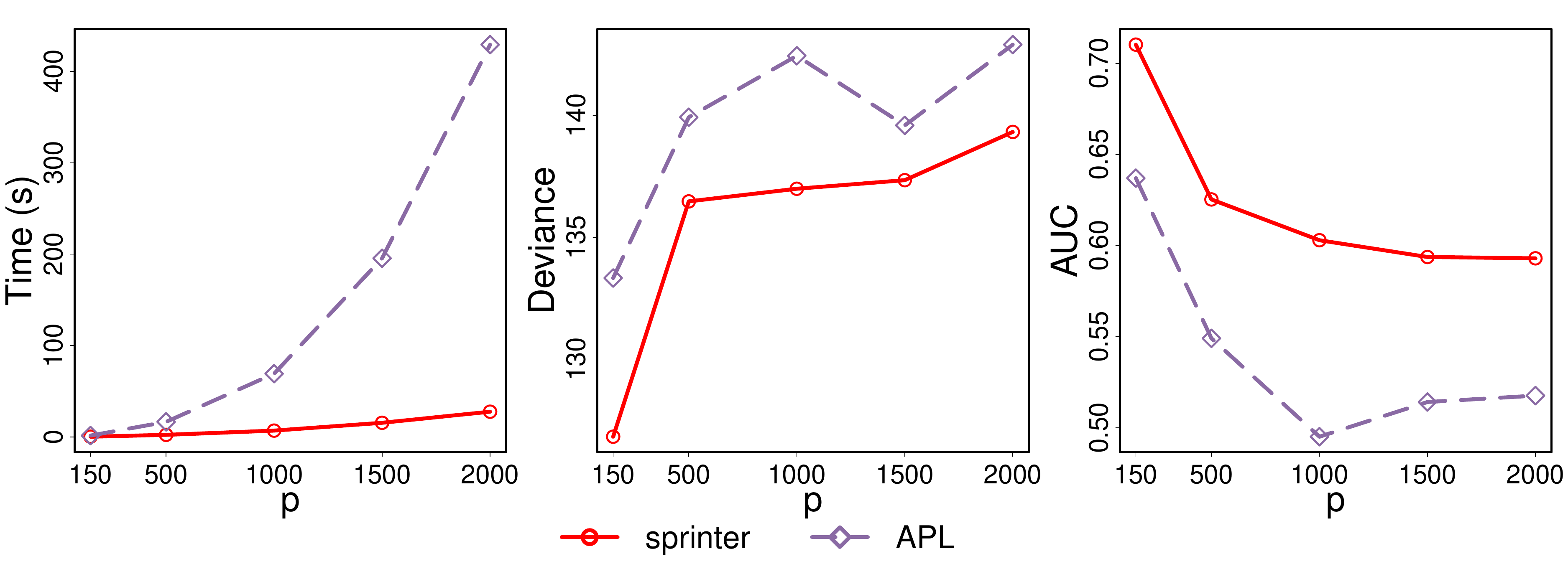}
    \caption{Logistic regression models: comparison of computation time (in seconds), deviance and AUC for different values of $p$ (averaged over 50 replications). The experiments are conducted on a 2020 Macbook Pro (13.3 inch, Apple M1 chip, RAM 8GB).}
  \label{fig:computation}
\end{center}
\end{figure}

Although in Section \ref{sec:sprinter} we illustrate that the worst-case complexity of \texttt{sprinter} is the same as that of \texttt{APL}, Figure \ref{fig:computation} shows that \texttt{sprinter} has obvious advantage in computational efficiency over \texttt{APL} in practice. Specifically, \texttt{sprinter} is much more scalable to large problem sizes. In particular, when $p=2000$ with around 2 million candidate interactions, \textit{sprinter} takes only 28 seconds (averaged over 50 repetitions), which is nearly 15 times faster than \texttt{APL}.

In addition to the favorable computational performance, \texttt{APL} also attains better predictive power as it is more effective in recruiting interactions.
Recall that \texttt{APL} treats main effects and interactions equally in the fitting so the overwhelmingly large number of candidate interaction will interfere \texttt{APL}'s ability to correctly model main effects.

\section{Application to the Tripadvisor hotel reviews}\label{sec:Tripadvisor}
We study the hotel reviews dataset from Tripadvisor \citep{wang2010latent}, where the goal is to build an interpretable predictive model of a hotel rating based on the words used in its reviews. This dataset has been studied in interaction modeling \citep{2019arXiv190708414Y} as well as other applications in high-dimensional statistics \citep{yan2021rare}.
The dataset contains a total of $n=211321$ reviews, each associated with a rating that takes integer value on the scale of $1$ to $5$ which we take as the response, and a text review. The text reviews were further processed to create a dictionary of $p = 7817$ different words, the majority of which are adjectives with some semantically transitioning words such as \emph{not}, \emph{but}, etc. 
By doing so, for each observation of hotel rating and review pair, we generate $p = 7817$ main effects, each of which is a binary predictor indicating if a specific word in the dictionary is contained in a review. 
By this definition of main effects, an interaction is the indicator that the two constituent words coexist in the review. 

\citet{2019arXiv190708414Y} study their proposed interaction modeling procedure, which as illustrated in Section \ref{sec:linear_app} is a special case of Algorithm \ref{alg:sprinter} in Gaussian linear models with interactions, in this data application. The demonstrated favorable predictive performance, when compared with \texttt{MEL} and \texttt{APL}, shows the importance of effectively including interactions for the prediction tasks in this application; and when compared with hierarchical methods such as \citet{hao2018model}, indicates the importance of developing interaction modeling framework that is free of the hierarchical assumptions.

However, \citet{2019arXiv190708414Y} ignores the crucial fact that the responses in this dataset take ordinal categorical values, e.g., a response value of $5$ represents a high rating and an $1$ represents a low rating. Both the categorical and ordinal nature of the responses are not captured by the linear model (with interactions) that is assumed in \citet{2019arXiv190708414Y}. In this section, we apply the proposed \texttt{sprinter} algorithm in \textit{proportional odds models} \citep{mccullagh1980regression}, which is a commonly used GLM for ordinal responses. Specifically, we consider
\begin{align}
    \operatorname{logit}[\operatorname{Pr}(Y \leq k)]=\log \left[\frac{P(Y \leq k)}{P(Y >k)}\right]=\alpha^\ast_{k} -\X^T \bbeta^\ast - \Z^T \bgamma^\ast, \quad \text{for} \,\, k \in\{1,2,3,4\},
    \label{mod:ordinal}
\end{align}
where the intercept $\alpha^\ast_k$ is specific to the response level $k$, and the coefficients $\bbeta^\ast$ for main effects and $\bgamma^\ast$ for interactions are shared across different response levels. 

In the implementation of Algorithm \ref{alg:sprinter} in this data application, we employ \texttt{ordinalNet} \citep{wurm2017regularized} for Step 1 and Step 4 with an $\ell_1$-penalty in both Step 1 and Step 4. We adapt the same strategy for tuning parameters selection as in Section \ref{sec:Time}.
We randomly select 10\% of the data as the training set, and the rest 90\% of the data as the test test.
For comparison, we round the predicted response values from \citet{2019arXiv190708414Y} to their nearest integers among 1 and 5. 
\begin{figure}[H]
\begin{center}
  \includegraphics[width=0.7\linewidth]{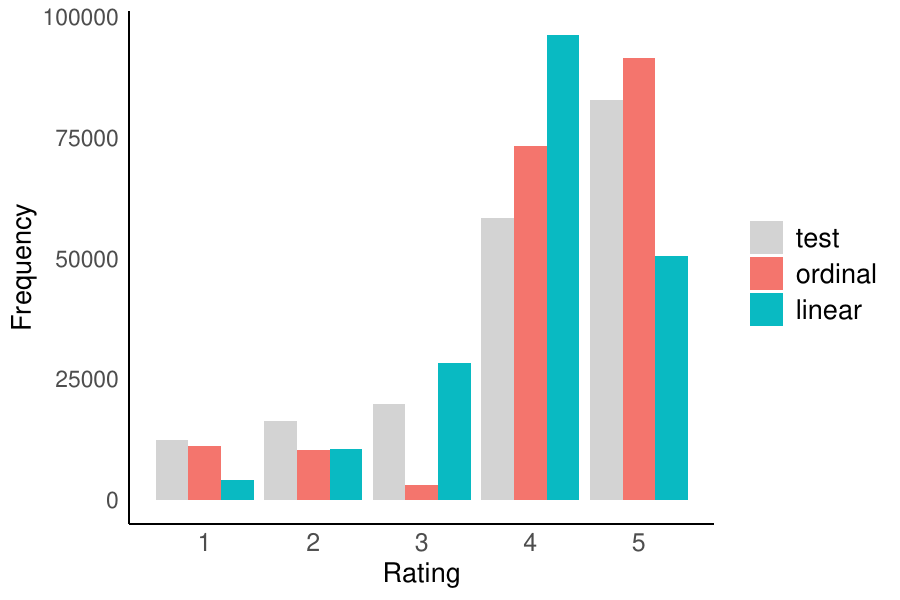}
  \caption{Histogram of the true ratings on the testing set (\texttt{test}), as well as their predicted values by \texttt{sprinter} in linear model with interactions \citep[\texttt{linear}, ][]{2019arXiv190708414Y} and proportional odds model (\texttt{ordinal}). }
  \label{fig:rating distribution}
\end{center}
\end{figure}
We observe, from Figure \ref{fig:rating distribution}, that the overall distribution of response is left-skewed, i.e., there are much more higher ratings. Compared with \citet{2019arXiv190708414Y}, \texttt{sprinter} with proportional odds model yields much better prediction, especially for higher ratings. Numerically, we obtain a $54.3\%$ prediction accuracy of \texttt{sprinter} with proportional odds model, which is much higher than $48.3\%$ of \texttt{sprinter} with Gaussian linear models as in \citet{2019arXiv190708414Y}. This observation illustrates the importance of developing an interaction modeling framework that is applicable to a wide range of models.

%\begin{table}[h!]
%\centering
%\begin{tabular}{cc}
%\hline \text{Method} & {Prediction accuracy}\\
%\hline \text{sprinter under linear regression} & 0.483 \\
% \text{sprinter under ordinal logistic regression} & 0.543 \\
%\hline 
%\end{tabular}
%\caption{prediction accuracy on the testing set}
%\label{tab:accu}
%\end{table}

In addition to the favorable predictive performance, the existence (main effects) and the co-existence (interactions) of the words selected by \texttt{sprinter} with \eqref{mod:ordinal} are highly interpretable, and they preserve interesting and important semantic meanings for implying hotels' business practices and strategies that could potentially increase their ratings. 

Firstly, we observe from \eqref{mod:ordinal} that, with everything else fixed, a positive coefficient in $\bbeta^\ast$ (or in $\bgamma^\ast$) indicates that the existence (or coexistence) of the corresponding word (or words) has the effect of increasing the probability that the response is in higher levels. In other words, the existence of words (and their interactions) that are believed to boost the ratings are expected to be associated with positive coefficients.

\begin{figure}[ht]
\begin{center}
  \includegraphics[width=0.7\linewidth]{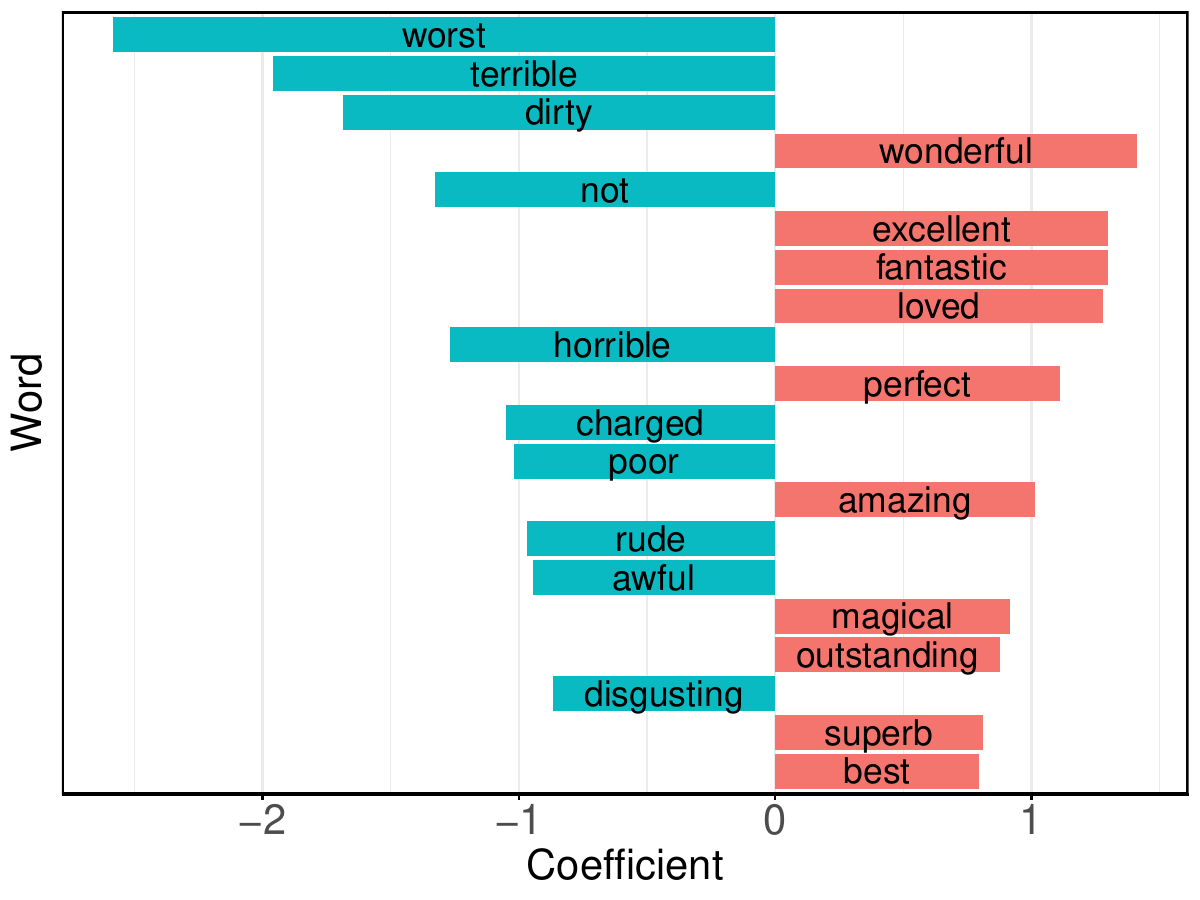}
  \caption{Top 10 recovered positive and negative main effects.}
  \label{fig:words main effects}
\end{center}
\end{figure}

As expected, the main effects selected by \texttt{sprinter} with \eqref{mod:ordinal}, as summarized in Figure \ref{fig:words main effects}, indeed meet this interpretation. Specifically, the selected main effects that have positive coefficients correspond to words that are semantically positive, such as \emph{wonderful} and \emph{excellent}, while words that correspond to negative coefficients have negative semantic meanings, such as \emph{worst} and \emph{terrible}. 

\begin{figure}[ht]
\begin{center}
  \includegraphics[width=0.7\linewidth]{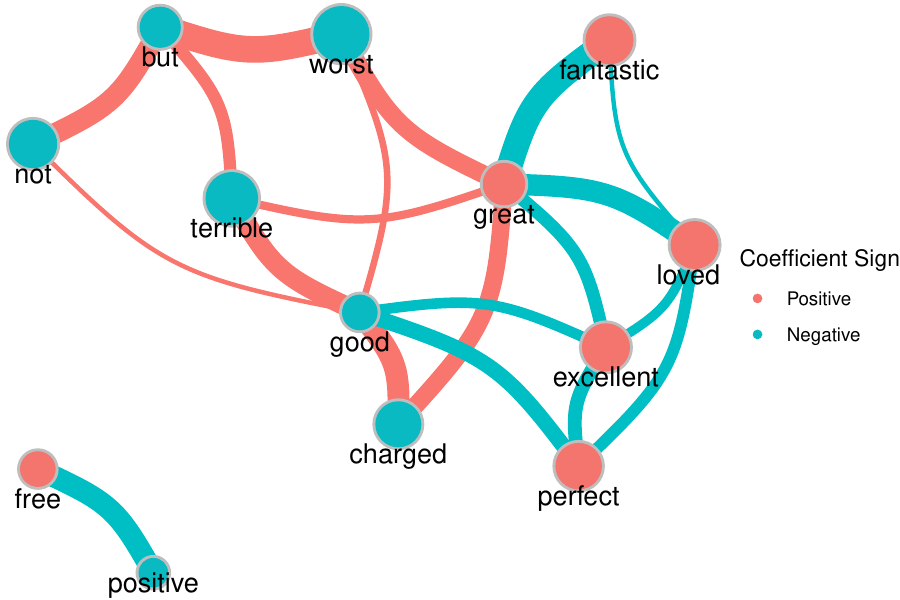}
  \caption{Top interactions estimation network. Each node represents a word, with its size indicating the coefficient's magnitude. Interactions between words are shown as edges, and their width signifies the interaction coefficients magnitude. Red color indicates positive coefficients, while blue color represents negative coefficients.}
  \label{fig:words interaction}
\end{center}
\end{figure}

We would expect similar interpretation of the selected interactions, which are shown in Figure \ref{fig:words interaction}. Surprisingly, however, we found that the coexistence of two semantically positive words, e.g., \emph{great} and \emph{fantastic}, usually tend to have a negative coefficient for the corresponding interaction and thus have a decreasing effect on the rating. This diminishing phenomenon from the superposition of synonyms is also observed in \citet{2019arXiv190708414Y}. Due to the presence-absence coding of the main effects, the interaction serves as a correction for the fact that the coexistence of words such as \emph{great} and \emph{fantastic} was treated the same as if the review contains the word \emph{great} twice in a main-effects-only model.

Furthermore, in interpreting selected interactions such as \emph{worst}$\times$\emph{but}, which surprisingly has a positive coefficient (and thus a positively semantic meaning), we note the effect of transition words, such as \emph{but}, in negating the semantic effect of its corresponding main effect in the rating. As an example, a review text says ``\emph{... we were expecting for the worst but...}''.

Finally, understanding the recovered interactions allows us to reexamine some of the main effects, particularly those that are supposed to be semantically neutral. Words such as \emph{positive} and \emph{good} have an interestingly negative coefficients: their existence in the review lead to lower ratings.

\section{Conclusion}
In this paper, we have introduced \texttt{sprinter}, a new framework for interaction modeling that is free of hierarchical assumptions and is applicable to a wide range of GLMs.  The proposed method adapts the reluctant interaction selection principle \citep{2019arXiv190708414Y}, but is developed in a conditional screening framework for interaction selection that does not depend on particular notions of residuals. Theoretically, we prove that the proposed method simultaneously recovers the set of pure interactions without resorting to heavy computation. Numerically, through simulated datasets and a real data application, we demonstrate the advantages of \texttt{sprinter} in predictive performance, computational efficiency, interpretability in the selected models, and the wide range of GLMs that \texttt{sprinter} is able to handle. 
The implementation details for the \texttt{sprinter} algorithm and the code that reproduces all of the numerical studies are available in the first author's \href{https://github.com/hluscience/reproducible/tree/main/sprintr}{GitHub Repository}.

The reluctant nature of interaction modeling in \texttt{sprinter} facilitates its potential to model interactions that is beyond pairwise interactions. In selecting higher-order interactions, one could effectively prioritize lower-order (over higher-order) interactions as we prioritize main effects (over interactions) as in this paper. We also note that while main effects are used in Step 1 in Algorithm \ref{alg:sprinter} for the simplicity of presentation, one could effectively use any $O(p)$ predictors derived from main effects, e.g., GAMs or trees. Theoretical understanding of these extensions are worthy of further investigations.

\bibliographystyle{agsm}
\bibliography{short.bib}

\newpage
\appendix
\section*{Appendices}
The main proof of Theorems are presented in Appendix \ref{app:main_proof}, and related lemma are shown in Appendix \ref{app:lemma}. In Appendix \ref{app:k_n and K_n}, we show the choice of $k_n$ and the dimensionality that the method can handle. The technical details of Section \ref{sec:linear_app}, which apply Theorem \ref{thm:equivalence} to linear models, are provided in Appendix \ref{app:linear_app_proof}. The screening property of ``top-$m$" approach is shown in Appendix \ref{app:topm}. We illustrate the computational tricks for improving the efficiency of the OrdinalNet Package in Appendix \ref{app:ordinet}. The additional results for Tripadvisor dataset application are shown in Appendix \ref{app:tripresult}.

\section{Main proofs} \label{app:main_proof}
\subsection{Proof of Theorem \ref{thm:equivalence}} 
\label{app:equivalence}
First, we derive some critical expressions for $\check{\gamma}_j$, $\gamma^M_j$ and $\Cov_L (Y, Z_j | \X)$ as stated in the theorem.

By the optimality condition for \eqref{eq:marginal} with respect to $(\check{\bbeta}, \check{\gamma}_j)$, we obtain
\begin{align}
  \E \left[ b' \left( \X^T \check{\bbeta} + Z_j \check{\gamma}_j \right) 
    \begin{pmatrix}
      \X \\
      Z_j
  \end{pmatrix} \right] 
  = \E \left[Y 
    \begin{pmatrix}
      \X \\
      Z_j
  \end{pmatrix} \right].
  \label{eq:score_marginal}
\end{align}
Similarly, the optimality condition for \eqref{eq:baseline} with respect to $\bbeta^M$ leads to
\begin{align}
  \E \left[ b' \left( \X^T \bbeta^M \right)  \X\right]  = \E[Y \X],
  \label{eq:score_baseline}
\end{align}
and for \eqref{eq:population} with respect to $\gamma^M_j$
\begin{align}
  \E \left[ b' \left( \X^T \bbeta^M + Z_j \gamma^M_j \right) Z_j\right]  = \E[Y Z_j].
  \label{eq:score_population}
\end{align}

Recall the conditional expectation of $Y$ on $\X$ is defined as
\begin{align}
  \E_L \left( Y | \X \right) = b'(\X^T \bbeta^M),
  \label{def:cle_Y}
\end{align}
and the conditional expectation of $Z_j$ on $\X$ is 
\begin{align}
  \qquad \E_L(Z_j | \X) = \Cov(Z_j, \X)\Cov(\X)^{-1} \X.
  \label{def:cle_Z}
\end{align}
Using \eqref{eq:score_baseline} and \eqref{def:cle_Y}, we have 
\begin{align}
  \E \left[ Y \X - \E_L(Y | \X) \X \right] = \0.
  \label{eq:cle_property}
\end{align}
Then, from \eqref{def:cle_Z} and \eqref{eq:cle_property}, the $\Cov_L (Y, Z_j | \X)$ in statement \texttt{S3} is
\begin{align}
  \Cov_L (Y, Z_j | \X) :&= \E \left\{ \left[ Z_j - \E_L (Z_j | \X) \right] \left[ Y - \E_L(Y | \X) \right] \right\} \nonumber\\
                        &= \E  Z_j \left[ Y - \E_L(Y | \X) \right]  - \Cov(Z_j, \X)\Cov(\X)^{-1} \E\left\{ \X \left[Y - \E_L(Y | \X) \right]\right\} \nonumber \\
                        &= \E  Z_j \left[ Y - \E_L(Y | \X) \right].
                        \label{eq:lincov}
\end{align}

We begin by showing that statement \texttt{S1} and statement \texttt{S3} are equivalent. If statement \texttt{S1} holds, i.e., $\check{\gamma}_j = 0$, by comparing \eqref{eq:score_marginal} and \eqref{eq:score_baseline}, we find that $\check{\bbeta} = \bbeta^M$ by Condition \ref{cond:unique}. Then, \eqref{eq:score_marginal} on the component $Z_j$ becomes
\begin{align}
   \E[YZ_j] = \E \left[ b'(\X^T \check{\bbeta}) Z_j\right]  = \E \left[ b'(\X^T \bbeta^M ) Z_j\right]  = \E \left[ \E_L \left( Y | \X \right) Z_j \right].
  \nonumber
\end{align}
Therefore, \eqref{eq:lincov} implies that $\Cov_L (Y, Z_j | \X) = 0$. This proves that statement \texttt{S1} implies statement \texttt{S3}.

To show statement \texttt{S3} implies statement \texttt{S1}, we note from \eqref{def:cle_Y} and \eqref{eq:lincov} that statement \texttt{S3} is equivalent to
\begin{align}
  \E \left[ b' (\X^T \bbeta^M) Z_j\right] = \E[Y Z_j].
  \label{eq:lincov_simple}
\end{align}
This, together with \eqref{eq:score_baseline}, implies that $(\bbeta^M, 0)^T$ is a solution to \eqref{eq:marginal}. By the uniqueness of the solution, we have $\check{\gamma}_j = 0$, therefore statement \texttt{S1} is proved.

Next we show the equivalence between statement \texttt{S2} and statement \texttt{S3}. If statement \texttt{S2} holds, i.e., $\gamma^M_j = 0$, then \eqref{eq:score_population} implies that
\begin{align}
  \E \left[ Y Z_j - \E_L \left( Y | \X \right) Z_j \right] = 0
  \nonumber
\end{align}
which is equivalent to statement \texttt{S3} by \eqref{eq:lincov}. If statement \texttt{S3} holds, then its equivalent equation \eqref{eq:lincov_simple} results in $\gamma^M_j = 0$ due to the unique solution to \eqref{eq:population}.

\subsection{Proof of Theorem \ref{thm:important}}
\label{app:important}
From \eqref{eq:lincov}, we have
  \begin{align}
    \left| \Cov_L \left( Y, Z_j | \X \right) \right| 
    &= \left |\E  Z_j \left[ Y - \E_L(Y | \X) \right]  \right|
    = \left| \E \left[ Z_j b' \left( \X^T \bbeta^M + Z_j \gamma^M_j \right) - Z_j b' \left( \X^T \bbeta^M \right) \right] \right| \nonumber\\
    &= \left| \gamma^M_j \E \left[ m_j Z_j^2 \right]\right| \leq c^{-1} \left| \gamma^M_j \right|, \nonumber
  \end{align}
  where the second equality holds from \eqref{eq:score_population} and \eqref{def:cle_Y}, and the inequality holds from Condition \ref{condition:b_prime}. Therefore, by the definition in \eqref{def:set}
  \begin{align}
    \min_{j \in \A(\kappa)}|\gamma^M_j| \geq c \min_{j \in \A(\kappa)} \left| \Cov_L \left( Y, Z_j | \X \right) \right| \geq c n^{-\kappa}.
    \nonumber
  \end{align}
  
\subsection{A characterization of noise level}
\label{app:concentration}
\begin{thm} \label{thm:concentration}
  Suppose that Condition \ref{condition:tail} to Condition \ref{condition:lowerbound} hold, and assume (without loss of generality) that $\E[X_j^4] \leq 4$ for all $j \in [p]$.  For any $t > 0$, take
  \begin{align}
    N = \frac{8}{V} (1 + t) \max \left\{ 4k_n n^{-1/2}, \left(4\tau^2  + \tau V\right)^{1/2} \EE \right\},
    \label{eq:N}
  \end{align}
  where 
  $\EE = \sqrt{\E [(\X^T ( \bbeta^M - \hat{\bbeta}))^2 ]}$ =  $\Theta(n^{-\rho})$ as in \eqref{eq:rho}. Then we have
  \begin{align}
    \Prob \left[ \max_{j \in [q]} |\hat{\gamma}_j - \gamma_j^M| \geq N\right] 
    \leq q \exp \left( -2 t^2 / K_n^2\right) + n q \left[s_1 \exp (-r_0 K_n^\alpha) + 2 r_1 \exp (-r_0 K_n^{\alpha / 2})\right].
    \nonumber
  \end{align}
\end{thm}
\begin{proof}
  Let, for any $j \in [q]$ and for any $N > 0$,
  \begin{align}
    \B_j(N) = \left\{ \gamma_j \in \B: |\gamma_j - \gamma^M_j| \leq N \right\}.
    \label{eq:Bj}
  \end{align}
  First we define a convex combination $\gamma^s_j = s \hat{\gamma}_j + (1 - s) \gamma^M_j$, where 
  \begin{align}
    s = \left(1 + \frac{|\hat{\gamma}_j - \gamma^M_j|}{N}\right)^{-1} \in (0, 1)
    \nonumber
  \end{align}
  for any $N > 0$. Then we have $\gamma^s_j \in \B_j(N)$. Convexity of $\ell$ implies that
  %\begin{align}
  %  \P_n \ell \left( \X^T \bbeta^M + Z_j \gamma^s_j, Y \right) \leq s \P_n \ell \left( \X^T \bbeta^M + Z_j \hat{\gamma}_j, Y \right) + (1 - s) \P_n \ell \left( \X^T \bbeta^M + Z_j \gamma^M_j, Y \right),
  %  \label{eq:convex}
  %\end{align}
  %and similarly
  \begin{align}
    \P_n \ell \left( \X^T \hat{\bbeta} + Z_j \gamma^s_j, Y \right) \leq s \P_n \ell \left( \X^T \hat{\bbeta} + Z_j \hat{\gamma}_j, Y \right) + (1 - s) \P_n \ell \left( \X^T \hat{\bbeta} + Z_j \gamma^M_j, Y \right),
    \nonumber
  \end{align}
  which further implies 
  \begin{align}
  &\P_n  \ell \left( \X^T \hat{\bbeta} + Z_j \gamma^s_j, Y \right) - \P_n \ell \left( \X^T \hat{\bbeta} + Z_j \gamma^M_j, Y \right) \nonumber \\
    \leq &s \P_n  \ell \left( \X^T \hat{\bbeta} + Z_j \hat{\gamma}_j, Y \right) - s \P_n \ell \left( \X^T \hat{\bbeta} + Z_j \gamma^M_j, Y \right) \leq 0,
    \label{eq:convex}
  \end{align}
  where the last inequality holds from the definition of $\hat{\gamma}_j$.

By the definition of $\gamma^M_j$, we have
  \begin{align}
    0 \leq & \E \left[ \ell \left( \X^T \bbeta^M + Z_j \gamma^s_j, Y \right) - \ell \left( \X^T \bbeta^M + Z_j \gamma^M_j, Y \right) \right] \nonumber\\
    =& (\E - \P_n) \left[ \ell \left( \X^T \bbeta^M + Z_j \gamma^s_j, Y \right) - \ell \left( \X^T \bbeta^M + Z_j \gamma^M_j, Y \right) \right]  \nonumber\\
     & + \P_n \left[ \ell \left( \X^T \bbeta^M + Z_j \gamma^s_j, Y \right) - \ell \left( \X^T \hat{\bbeta} + Z_j \gamma^s_j, Y \right) \right] \nonumber\\
     & + \P_n \left[ \ell \left( \X^T \hat{\bbeta} + Z_j \gamma^s_j, Y \right) - \ell \left( \X^T \hat{\bbeta} + Z_j \gamma^M_j, Y \right) \right] \nonumber\\
     & + \P_n \left[ \ell \left( \X^T \hat{\bbeta} + Z_j \gamma^M_j, Y \right) - \ell \left( \X^T \bbeta^M + Z_j \gamma^M_j, Y \right) \right] \nonumber\\
    \leq &(\E - \P_n) \left[ \ell \left( \X^T \bbeta^M + Z_j \gamma^s_j, Y \right) - \ell \left( \X^T \bbeta^M + Z_j \gamma^M_j, Y \right) \right]  \nonumber\\
         & + \P_n \left[ \ell \left( \X^T \bbeta^M + Z_j \gamma^s_j, Y \right) - \ell \left( \X^T \hat{\bbeta} + Z_j \gamma^s_j, Y \right) \right] \nonumber\\
         & + \P_n \left[ \ell \left( \X^T \hat{\bbeta} + Z_j \gamma^M_j, Y \right) - \ell \left( \X^T \bbeta^M + Z_j \gamma^M_j, Y \right) \right] \nonumber\\
    = &\E \left[ \ell \left( \X^T \bbeta^M + Z_j \gamma^s_j, Y \right) - \ell \left( \X^T \bbeta^M + Z_j \gamma^M_j, Y \right) \right]  \nonumber\\
      & + \P_n \left[ \ell \left( \X^T \hat{\bbeta} + Z_j \gamma^M_j, Y \right) - \ell \left( \X^T \hat{\bbeta} + Z_j \gamma^s_j, Y \right) \right] \nonumber\\
    = &\E \left[ \ell \left( \X^T \bbeta^M + Z_j \gamma^s_j, Y \right) - \ell \left( \X^T \bbeta^M + Z_j \gamma^M_j, Y \right) \right]  \nonumber\\
      & + \E \left[ \ell \left( \X^T \hat{\bbeta} + Z_j \gamma^M_j, Y \right) - \ell \left( \X^T \hat{\bbeta} + Z_j \gamma^s_j, Y \right) \right] \nonumber\\
      & + (\P_n - \E) \left[ \ell \left( \X^T \hat{\bbeta} + Z_j \gamma^M_j, Y \right) - \ell \left( \X^T \hat{\bbeta} + Z_j \gamma^s_j, Y \right) \right] \nonumber\\
    = &\E \left[ b \left( \X^T \bbeta^M + Z_j \gamma^s_j\right) - b \left( \X^T \bbeta^M + Z_j \gamma^M_j \right) \right]  \nonumber\\
      & + \E \left[ b \left( \X^T \hat{\bbeta} + Z_j \gamma^M_j \right) - b \left( \X^T \hat{\bbeta} + Z_j \gamma^s_j \right) \right] \nonumber\\
      & +  (\P_n - \E) \left[ \ell \left( \X^T \hat{\bbeta} + Z_j \gamma^M_j, Y \right) - \ell \left( \X^T \hat{\bbeta} + Z_j \gamma^s_j, Y \right) \right] \label{eq:basic}.
  \end{align}
  where the second inequality holds from \eqref{eq:convex}, the last equality holds from the fact that $\ell(\theta, Y) = b(\theta) - \theta Y$. 
  
  Furthermore, for any $j \in [q]$ and any $\gamma_j \in \B_j(N)$, 
  \begin{align}
  & \left| b \left( \X^T \bbeta^M + Z_j \gamma_j \right) - b \left( \X^T \hat{\bbeta} + Z_j \gamma_j \right) + b \left( \X^T \hat{\bbeta} + Z_j \gamma^M_j \right) - b \left( \X^T \bbeta^M + Z_j \gamma^M_j \right) \right| \nonumber\\
  = & \left| b \left( \X^T \bbeta^M + Z_j \gamma_j \right) - b \left( \X^T \hat{\bbeta} + Z_j \gamma_j \right)+ b \left( \X^T \hat{\bbeta} + Z_j \gamma^M_j \right) - b \left( \X^T \bbeta^M + Z_j \gamma^M_j \right) \right. \nonumber\\
    & \left. - \left[ b'(\X^T \hat{\bbeta} + Z_j \gamma_j) - b'(\X^T \hat{\bbeta} + Z_j \gamma_j) \right] \X^T(\bbeta^M - \hat{\bbeta}) \right. \nonumber\\
    & \left.  - \left[ b'(\X^T \bbeta^M + Z_j \gamma^M_j) - b'(\X^T \bbeta^M + Z_j \gamma^M_j) \right] \X^T (\hat{\bbeta} - \bbeta^M) \right| \nonumber\\
  \leq & \left| b \left( \X^T \bbeta^M + Z_j \gamma_j \right) - b \left( \X^T \hat{\bbeta} + Z_j \gamma_j \right) - b'(\X^T \hat{\bbeta} + Z_j \gamma_j) \X^T (\bbeta^M - \hat{\bbeta}) \right| \nonumber\\
       & + \left| b \left( \X^T \hat{\bbeta} + Z_j \gamma^M_j \right) - b \left( \X^T \bbeta^M + Z_j \gamma^M_j \right) - b'(\X^T \bbeta^M + Z_j \gamma^M_j) \X^T (\hat{\bbeta} - \bbeta^M) \right| \nonumber\\
       & + \left| b'(\X^T \hat{\bbeta} + Z_j \gamma_j) - b'(\X^T \bbeta^M + Z_j \gamma^M_j) \right| \left| \X^T (\hat{\bbeta} - \bbeta^M) \right| \nonumber\\
  \leq & \left| b'( \X^T \hat{\bbeta} + Z_j \gamma_j ) - b'( \X^T \bbeta^M + Z_j \gamma^M_j ) \right| \left| \X^T ( \bbeta^M - \hat{\bbeta}) \right| + \tau ( \X^T (\bbeta^M - \hat{\bbeta}) )^2 \nonumber\\
  \leq & \left|b'( \X^T \hat{\bbeta} + Z_j \gamma_j ) - b'( \X^T \hat{\bbeta} + Z_j \gamma^M_j ) \right| \left|\X^T ( \bbeta^M - \hat{\bbeta})\right| \nonumber\\
       & +  \left| b'( \X^T \hat{\bbeta} + Z_j \gamma^M_j ) - b'( \X^T \bbeta^M + Z_j \gamma^M_j ) \right| \left|\X^T ( \bbeta^M - \hat{\bbeta}) \right| + \tau ( \X^T (\bbeta^M - \hat{\bbeta}) )^2 \nonumber\\
  \leq & \tau |Z_j (\gamma_j - \gamma^M_j)|\left|\X^T ( \bbeta^M - \hat{\bbeta}) \right| + 2\tau \left(\X^T ( \bbeta^M - \hat{\bbeta}) \right)^2, \nonumber
\end{align}
where the second inequality comes from Lemma \ref{lem:tausmooth}, and the fourth inequality comes directly from Condition \ref{condition:tausmooth}.  Therefore, from \eqref{eq:basic} we have
\begin{align}
  0 \leq & \E \left[ \ell \left( \X^T \bbeta^M + Z_j \gamma^s_j, Y \right) - \ell \left( \X^T \bbeta^M + Z_j \gamma^M_j, Y \right) \right] \nonumber\\
  \leq & \sup_{\gamma_j \in \B_j(N)} \left| (\P_n - \E) \left[ \ell \left( \X^T \hat{\bbeta} + Z_j \gamma^M_j, Y \right) - \ell \left( \X^T \hat{\bbeta} + Z_j \gamma_j, Y \right) \right] \right| \nonumber \\
       & + \sup_{\gamma_j \in \B_j(N)} \tau \E \left[ \left|Z_j (\gamma_j - \gamma^M_j) \X^T (\bbeta^M - \hat{\bbeta}) \right| \right] + 2 \tau \E \left[\left(\X^T ( \bbeta^M - \hat{\bbeta}) \right)^2 \right] \nonumber\\
  \leq & \sup_{\gamma_j \in \B_j(N)} \left| (\P_n - \E) \left[ \ell \left( \X^T \hat{\bbeta} + Z_j \gamma^M_j, Y \right) - \ell \left( \X^T \hat{\bbeta} + Z_j \gamma_j, Y \right) \right] \right| \nonumber \\
       & +  \tau N \max_{j \in [q]} \E \left[ \left|Z_j\X^T (\bbeta^M - \hat{\bbeta}) \right| \right] + 2 \tau \E \left[\left(\X^T ( \bbeta^M - \hat{\bbeta}) \right)^2 \right] \nonumber \\
  \leq & G(N) +  2\tau N  \EE + 2 \tau \EE^2, \nonumber
\end{align}
where
\begin{align}
  G(N) = \sup_{\gamma_j \in \B_j(N)} \left|(\P_n - \E) \left[ \ell \left( \X^T \hat{\bbeta} + Z_j \gamma^M_j, Y \right) - \ell \left( \X^T \hat{\bbeta} + Z_j \gamma_j, Y \right) \right] \right| \label{eq:GN}.
\end{align}
By Condition \ref{condition:lowerbound}, we have
\begin{align}
  \left| \gamma^s_j - \gamma^M_j \right| \leq \left[ \frac{G(N) + 2 \tau N  \EE + 2 \tau \EE^2}{V} \right]^{1/2}.
  \nonumber
\end{align}
So for any $x > 0$,
\begin{align}
  \Prob \left( \left| \gamma^s_j - \gamma^M_j \right| \geq x \right) \leq \Prob \left( G(N) + 2 \tau N \EE + 2 \tau \EE^2 \geq V x^2 \right).
  \nonumber
\end{align}
Setting $x = N / 2$ and using the definition of $\gamma^s_j$, we have
\begin{align}
  \Prob \left( \left| \hat{\gamma}_j - \gamma^M_j \right| \geq N \right) 
  = \Prob \left( \left| \gamma^s_j - \gamma^M_j \right| \geq N / 2 \right)
  \leq \Prob \left( G(N) + 2 \tau N \EE + 2 \tau \EE^2 \geq V N^2 / 4 \right). \nonumber
\end{align}

Simple algebra shows that for any
\begin{align}
  N \geq 4 V^{-1} \left[ 2 \tau + \sqrt{4 \tau^2  + \tau V} \right]\EE,
  \nonumber
\end{align}
we have
\begin{align}
  V N^2 /4 - 2 \tau N \EE - 2 \tau \EE^2 \geq V N^2 / 8. \nonumber
\end{align}

Now for any $t > 0$, take $N$ as in \eqref{eq:N}, we have
\begin{align}
  \Prob \left( \left| \hat{\gamma}_j - \gamma^M_j \right| \geq N \right) 
  &\leq \Prob \left( G(N) + 2 \tau N \EE + 2 \tau \EE^2 \geq V N^2 / 4 \right) \nonumber \\
  &\leq \Prob \left( G(N) \geq V N^2 / 8 \right) \nonumber\\
  &\leq \Prob \left( G(N) \geq  4 N k_n n^{-1/2} (1 + t) \right) \nonumber\\
  &\leq \Prob \left( G(N) \geq 4 N k_n n^{-1/2} (1 + t), \indi_{n,j} = 1 \right) + \Prob \left( \indi_{n,j} = 0 \right), \label{eq:tailbound}
\end{align}
where for $\Lambda_j$ in \eqref{eq:lambdaj}, we define
\begin{align}
  \indi_{n, j} := \prod_{i = 1}^n \indi\left\{ (z_{ij}, y_i) \in \Lambda_j \right\}.
  \label{eq:indi}
\end{align}

The rest of this proof focuses on characterizing the right hand side of \eqref{eq:tailbound}. From \eqref{eq:GN},
\begin{align}
  G(N) =& \sup_{\gamma_j \in \B_j(N)} \left|(\P_n - \E) \left[ \ell \left( \X^T \hat{\bbeta} + Z_j \gamma^M_j, Y \right) - \ell \left( \X^T \hat{\bbeta} + Z_j \gamma_j, Y \right) \right] \left( \indi_{n,j} + 1 - \indi_{n,j} \right) \right| \nonumber\\
  \leq &\sup_{\gamma_j \in \B_j(N)} \left|(\P_n - \E) \left[ \ell \left( \X^T \hat{\bbeta} + Z_j \gamma^M_j, Y \right) - \ell \left( \X^T \hat{\bbeta} + Z_j \gamma_j, Y \right) \right] \indi_{n,j} \right| \nonumber\\
       &+ \sup_{\gamma_j \in \B_j(N)} \left|(\P_n - \E) \left[ \ell \left( \X^T \hat{\bbeta} + Z_j \gamma^M_j, Y \right) - \ell \left( \X^T \hat{\bbeta} + Z_j \gamma_j, Y \right) \right] \left(1 -\indi_{n,j} \right) \right|. \nonumber
\end{align}
Note that conditional on the event $\{\indi_{n,j} = 1\}$, we have
\begin{align}
  \sup_{\gamma_j \in \B_j(N)} \left|\P_n \left[ \ell \left( \X^T \hat{\bbeta} + Z_j \gamma^M_j, Y \right) - \ell \left( \X^T \hat{\bbeta} + Z_j \gamma_j, Y \right) \right] \left(1 -\indi_{n,j} \right) \right| = 0.
  \nonumber
\end{align}
Thus conditional on the event $\{\indi_{n,j} = 1\}$, Lemma \ref{lem:lip} implies that
\begin{align}
  G(N) \leq & G_1(N) + \sup_{\gamma_j \in \B_j(N)} \left|\E \left[ \ell \left( \X^T \hat{\bbeta} + Z_j \gamma^M_j, Y \right) - \ell \left( \X^T \hat{\bbeta} + Z_j \gamma_j, Y \right) \right] \left(1 -\indi_{n,j} \right) \right| \nonumber\\
  =& G_1(N) + o(n^{-1}),
  \nonumber
\end{align}
where
\begin{align}
  G_1(N) := \sup_{\gamma_j \in \B_j(N)} \left|(\P_n - \E) \left[ \ell \left( \X^T \hat{\bbeta} + Z_j \gamma^M_j, Y \right) - \ell \left( \X^T \hat{\bbeta} + Z_j \gamma_j, Y \right) \right] \indi_{n,j} \right|.  
  \nonumber
\end{align}
Then \eqref{eq:tailbound}, Lemma \ref{lem:GN1}, and Lemma \ref{lem:indi}  together imply that 
\begin{align}
  \Prob \left( \left| \hat{\gamma}_j - \gamma^M_j \right| \geq N \right) 
  &\leq \Prob \left( G_1(N) \geq 4 N k_n n^{-1/2} (1 + t) - o(n^{-1}), \indi_n(j) = 1\right) + \Prob \left( \indi_n(j) = 0 \right) \nonumber\\
  &\leq \exp \left( -2 t^2 / K_n^2\right) + n \left[s_1 \exp (-r_0 K_n^\alpha) + 2 r_1 \exp (-r_0 K_n^{\alpha / 2})\right], \nonumber
\end{align}
and then the Theorem \ref{thm:concentration} follows with a union bound over $j \in [q]$.

\end{proof}

\subsection{Proof of Theorem \ref{thm:screening}} \label{app:proof_screening}
For any $c_1 > 0$, take
\[
  1 + t = \frac{c_1 V}{8} n^{-\kappa} \min \left\{ \frac{1}{4k_n} n^{1/2}, \frac{\tilde{c}}{\left(4 \tau^2  + \tau V \right)^{1/2}}  n^{\rho} \right\}.
\]

Note that $t>0$ only when $\kappa < \min\{1/2, \rho\}$ for sufficiently large $n$. By plugging $(1+t)$ into \eqref{eq:N}, the statement \texttt{S1} is proved by Theorem \ref{thm:concentration}.

For statement \texttt{S2}, we note that for any $j \in \A(\kappa)$, we have
\begin{align}
  |\hat{\gamma}_j| = |\gamma^M_j + \hat{\gamma}_j - \gamma^M_j| 
  \geq |\gamma^M_j| - |\hat{\gamma}_j - \gamma^M_j|.
  \nonumber
\end{align}

For the constant $c$ in Condition \ref{condition:b_prime}, on the event $\mathcal{T}_j = \{|\hat{\gamma}_j - \gamma^M_j| < c n^{-\kappa} / 2\}$, Theorem \ref{thm:important} implies that:
\[
|\hat{\gamma}_j| \geq c n^{-\kappa} - \frac{c}{2} n^{-\kappa} = \frac{c}{2} n^{-\kappa} \geq c_3 n^{-\kappa},
\]
which implies that $j \in \hat{\I}_\eta$ with the choice of $\eta = c_3 n^{-\kappa}$ where $c_3 \leq c / 2$. Therefore, on the event $\mathcal{T}_j$, we have  $\A(\kappa) \subseteq \hat{\I}_\eta$. This leads to:
\begin{align}
  &1 - \Prob \left[ \A(\kappa) \subseteq \hat{\I}_\eta \right] = \Prob \left[\bigcup_{j \in \A(\kappa)} \mathcal{T}_j^C\right] \nonumber\\
  &\leq |\A(\kappa)| \left[ \exp \left( - \frac{c_2}{K_n^2} \min \left\{ k_n^{-2} n^{1 - 2 \kappa}, n^{2 \rho - 2 \kappa} \right\}\right) + n  \left[(r_1 + s_1) \exp (-r_0 K_n^\alpha) + 2 r_1 \exp (-r_0 K_n^{\alpha / 2})\right] \right]
  \nonumber
\end{align}

\subsection{Proof of Theorem \ref{thm:size}}
First, we build the connection between $\Cov_L (Y, Z_j | \X)$ and $\bSigma_{\Z | \X}$. Recall that without loss of generality, we assume that $\E[\X] = \0$, then 
$\E[\X\X^T] = \bSigma$, $\E(\Z\Z^T) = \bPsi$, and $\E [\X\Z^T] = \bPhi$, as introduced in Section \ref{sec:linear_app}. This gives us $\E_L(\Z | \X) = \bPhi \bSigma^{-1} \X$. Furthermore,
\begin{align}
  \bSigma_{\Z | \X} :=& \E\{ [\Z - E_{L}(\Z | \X)] [\Z - E_{L}(\Z | \X)]^T \} \nonumber\\
  =& \bPsi - \bPhi \bSigma^{-1} \bPhi.
  \label{eq:covzx}
\end{align}

\label{app:size}
Here, we use $\omega$ to denote the interaction signal that cannot be captured by main effects fitting
\begin{align}
  \omega = \X^T \bbeta^\ast + \Z^T \bgamma^\ast - \X^T \bbeta^M.
  \nonumber
\end{align}
It is easy to verify that
\begin{align}
  \E (\Z\omega) 
  =& \E \left[ \E_L[\Z | \X] \omega \right] + \E \left[ \left(\Z - \E_L[\Z | \X]\right) \omega \right] \nonumber\\
  =& \bnu + \left(\bPsi - \bPhi \bSigma^{-1} \bPhi \right) \bgamma^\ast \nonumber\\
  :=& \bnu + \bSigma_{\Z | \X} \bgamma^\ast,
  \nonumber
\end{align}
where $\bnu$ is defined as in \eqref{def:w}.

Then, from \eqref{eq:lincov}, we have
\begin{align}
  \left|\Cov_L (Y, Z_j | \X) \right| &= \left| \E Z_j \left[ Y - \E_L(Y | \X) \right] \right| \nonumber\\
                     &= \left| \E \left[ Z_j Y - Z_j b'(\X^T \bbeta^M) \right] \right| \nonumber\\
                     &= \left| \E \left[ Z_j b'(\X^T \bbeta^\ast + \Z^T \bgamma^\ast) - Z_j b'(\X^T \bbeta^M) \right] \right|  \nonumber\\
                     &\leq \tau \left| \E \left[ Z_j (\X^T \bbeta^\ast + \Z^T \bgamma^\ast - \X^T \bbeta^M) \right] \right|  \nonumber\\
                     &= \tau \left| \E \left[ Z_j \omega \right] \right|,  \nonumber
\end{align}
where the third equality comes from the fact $\E(Y | \X) = b'(\X^T \bbeta^\ast + \Z^T \bgamma^\ast)$ and the inequality holds from Condition \ref{condition:tausmooth}.
Thus
\begin{align}
  \sum_{j = 1}^q \Cov_L(Y, Z_j | \X)^2 
  &\leq \tau^2 \norm{\E \left(\Z \omega \right)}_2^2 = \tau^2 \norm{\bnu + \bSigma_{\Z|\X}\bgamma^\ast}_2^2 \nonumber\\
  & = \tau^2 \left\{ \norm{\bnu}_2^2 + 2 \bnu^T \bSigma_{\Z|\X} \bgamma^\ast + {\bgamma^\ast}^T \bSigma_{\Z|\X}^{1/2} \bSigma_{\Z | \X} \bSigma_{\Z | \X}^{1/2}\bgamma^\ast \right\} \nonumber\\
  & \leq \tau^2 \left\{ \norm{\bnu}_2^2 + 2 \bnu^T \bSigma_{\Z|\X} \bgamma^\ast + \lambda_{\max} \left( \bSigma_{\Z | \X} \right) {\bgamma^\ast}^T \bSigma_{\Z | \X}\bgamma^\ast\right\} \nonumber\\
  & \leq \tau^2 \left\{ \norm{\bnu}_2^2 + 2 \bnu^T \bSigma_{\Z|\X} \bgamma^\ast + \lambda_{\max} \left( \bSigma_{\Z | \X} \right) \Var(\Z^T \bgamma^\ast) \right\} \nonumber\\
  & = O \left( \lambda_{\max} \left( \bSigma_{\Z | \X} \right) \right),
  \label{eq:ub_sumcov}
\end{align}
where the first equality holds from \eqref{def:w}, the last equality holds because of Condition \ref{condition:size}, and the third inequality holds from the law of total variance and Condition \ref{condition:size}, i.e., 
\begin{align}
  {\bgamma^\ast}^T \bSigma_{\Z | \X}\bgamma^\ast
&= {\bgamma^\ast}^T \E \left\{ \left[ \Z - \E_L(\Z | \X) \right]\left[ \Z - \E_L(\Z | \X) \right]^T \right\} \bgamma^\ast \nonumber\\
&= \E \left[ \left( \Z^T \bgamma^\ast - \E_L (\Z^T \bgamma^\ast | \X) \right)^2\right] \nonumber\\
&= \E \left\{ \E \left[ \left( \Z^T \bgamma^\ast - \E_L (\Z^T \bgamma^\ast | \X) \right)^2 | \X\right] \right\} \nonumber\\
&= \E \left\{ \Var (\Z^T \bgamma^\ast | \X) \right\} \nonumber\\
&= \Var \left( \Z^T \bgamma^\ast \right) - \Var \left[ \E \left( \Z^T \bgamma^\ast | \X \right) \right] \nonumber\\
&\leq \Var(\Z^T \bgamma^\ast) = O(1).
\nonumber
\end{align}

Next, in order to connect the relationship between $\Cov_L (Y, Z_j | \X)$ and $\gamma_j^M$, recall from the definition of $m_j$ in Condition \ref{condition:b_prime}, we have that
\begin{align}
  \Cov_L \left( Y, Z_j | \X \right) 
  &= \E \left[ Z_j \left(Y - \E_L (Y | \X) \right)\right] \nonumber\\
  &= \E \left[ Z_j \left(b'\left( \X^T \bbeta^M + Z_j \gamma_j^M \right) - b'(\X^T \bbeta^M) \right) \right] \nonumber\\
  &= \gamma_j^M \E \left( m_j Z_j^2 \right),
  \nonumber
\end{align}
Part (2) of Condition \ref{condition:size} implies that
\begin{align}
  |\gamma_j^M| = O \left( \Cov_L \left( Y, Z_j | \X \right)\right),
  \nonumber
\end{align}
which together with \eqref{eq:ub_sumcov} implies that
\begin{align}
  \norm{\bgamma^M}_2^2 = \sum_{j=1}^{q} |\gamma_j^M|^2 = O( \sum_{j=1}^{q}  \Cov_L \left( Y, Z_j | \X \right)^2) = O \left( \lambda_{\max} \left(\bSigma_{\Z| \X}\right)\right).
  \nonumber
\end{align}
Now $\forall \varepsilon >0$,
we have
\[
  \left| \left\{ j \in [q]: \left| \hat{\gamma}_j \right| \geq 2 \varepsilon n^{-\kappa} \right\} \right| \leq 
  \left| \left\{ j \in [q]: \left| \gamma^M_j \right| \geq \varepsilon n^{-\kappa} \right\} \right|  = O \left( n^{2\kappa} \lambda_{\max}\left(\bSigma_{\Z | \X} \right)\right).
\]
The result then follows by taking $\varepsilon = c_3 / 2$ where $c_3$ is the constant in part 2 of Theorem \ref{thm:screening}.

\section{Technical lemma}\label{app:lemma}
\begin{lem}[Adapting Lemma 1 in \citet{fan2010sure}] \label{lem:taily}
  If Condition \ref{condition:tail} holds, than for any $t > 0$, 
  \begin{align}
    \Prob \left( \left| Y \right| \geq r_0 t^\alpha / s_0 \right) \leq s_1 \exp \left( -r_0 t^\alpha \right).
    \nonumber
  \end{align}
\end{lem}
\begin{proof}
  By Markov inequality,
  \begin{align}
    \Prob \left( Y \geq u \right) \leq \exp(-s_0 u) \E \left[ \exp(s_0 Y) \right].
    \nonumber
  \end{align}
  Since the distribution of $Y$ belongs to exponential family, we have
  \begin{align}
    \E \left[ \exp(s_0 Y) | \theta \right] = \exp \left[ b(\theta + s_0) - b(\theta) \right].
    \nonumber
  \end{align}
  
  Thus we have
  \begin{align}
    \Prob \left( Y \geq u \right) \leq \exp (-s_0 u) \E \left[\exp \left[ b(\theta + s_0) - b(\theta) \right] \right]
    \nonumber
  \end{align}
  and similarly, 
    \begin{align}
    \Prob \left( Y \leq -u \right) \leq \exp (-s_0 u) \E \left[\exp \left[ b(\theta - s_0) - b(\theta) \right] \right].
    \nonumber
  \end{align}

  By Condition \ref{condition:tail} with $\theta = \X^T \bbeta^\ast + \Z^T \bgamma^\ast$,
  \begin{align}
    \Prob \left( |Y| \geq u \right) \leq  s_1 \exp(-s_0 u)  
    \nonumber
  \end{align}
  and the proof completes by taking $u = r_0 t^\alpha / s_0$.
\end{proof}

\begin{lem} \label{lem:tausmooth}
  Under Condition \ref{condition:tausmooth}, we have for any $s, t$,
  \begin{align}
    \left| b(s) - b(t) - b'(t) (s - t) \right| \leq \frac{\tau}{2} (s - t)^2.
    \nonumber
  \end{align}
\end{lem}

\begin{proof}
  \begin{align}
  &\left| b(s) - b(t) - b'(t) (s - t) \right| \nonumber\\
    = &\left| \int_{0}^1 b'(t + u(s - t)) (s - t) du - b'(t) (s - t) \right| \nonumber\\ 
    = &\left| \int_{0}^1  \left[b'(t + u(s - t)) - b'(t)  \right] (s - t) du  \right| \nonumber\\ 
    \leq & \int_{0}^1  \left| b'(t + u(s - t)) - b'(t)  \right| |s - t| du \nonumber\\ 
    \leq & \int_{0}^1  \tau u |s - t|^2 du = \frac{\tau}{2}  (s - t)^2.  \nonumber
  \end{align}
\end{proof}

\begin{lem} \label{lem:lip}
  Under Condition \ref{condition:lip}, there exists a sufficiently large constant $C$ such that with $b_n$ and $V$ given in Condition \ref{condition:lowerbound}, for any $a_0 \in \B$,
  \begin{align}
    \sup_{\gamma_{j} \in \B_j(N)} \left|\E \left[ \ell \left(a_0 + Z_j\gamma_j, Y\right)-\ell \left(a_0 + Z_j\gamma_j^{M}, Y\right)\right]\right| \indi \left\{ (Z_j, Y) \notin \Lambda_j \right\} = o\left(n^{-1}\right). 
    \label{eq:littleon}
  \end{align}
\end{lem}
\begin{proof}
  For any $j \in [q]$,
    \begin{align}
    &\left|\E \left[ \ell \left(a_0 + Z_j\gamma_j, Y\right)-\ell \left(a_0 + Z_j\gamma_j^{M}, Y\right)\right]\indi \left\{ (Z_j, Y) \notin \Lambda_j \right\}\right|  \nonumber\\
    \leq&  \left| \E \left[ b(a_0 + Z_j \gamma_j)  \indi \left\{|Z_j| \geq K_n \right\} \right]\right| + \left| \E \left[ b(a_0 + Z_j \gamma^M_j) \indi \left\{ |Z_j| \geq K_n \right\}  \right]\right| \nonumber\\
        & +  \E \left[ \left| Y Z_j \left(\gamma_j - \gamma^M_j\right) \left(1 - \indi \left\{ |Z_j| \leq K_n, |Y| \leq K_n^\ast \right\}  \right)  \right]\right| \nonumber\\
    \leq& o(n^{-1}) + 2 N \E \left[ \left| Y Z_j \left(1 - \indi \left\{ |Z_j| \leq K_n, |Y| \leq K_n^\ast \right\}  \right)  \right]\right| \nonumber \\
    \leq& o(n^{-1}) + N \E \left[ \left| Y^2 \indi \left\{ |Y| > K_n^\ast \right\}  \right]\right| + N \E \left[ \left| Z_j^2 \indi \left\{ |Z_j| > K_n\right\} \right]\right| \nonumber\\
    =& o(n^{-1}) + N \int^\infty_{{K_n^\ast}^2} \Prob (Y^2 > t) dt  + N \int_{K^2_n}^\infty \Prob \left( Z_j^2 > t \right)dt \nonumber\\
    = & o(n^{-1}), \nonumber
  \end{align}
where the $N$ is defined \eqref{eq:N} and the second inequality follows from Condition \ref{condition:lip} and the last equality follows from the derivation below: 
  \begin{align}
  \int^\infty_{{K_n^\ast}^2} \Prob (Y^2 > t) dt 
  &\stackrel{t=u^2} = 2\int^\infty_{K_n^\ast} \Prob (|Y| > u) udu \nonumber \\
  & < 2\int^\infty_{K_n^\ast} s_1 \exp \left(-s_0 u\right) udu \nonumber \\
  & = \frac{2 K_n^* s_1}{s_0} \exp \left(-s_0 K_n^*\right)+\frac{2 s_1}{s_0^2} \exp \left(-s_0 K_n^*\right) \nonumber \\
  & < O(n^{-1}). \nonumber
  \end{align}
Here, the first inequality comes from the proof of Lemma \ref{lem:taily}, and the second equality is based on the integration by parts, and the last inequality holds due to the exponential decay being faster than the polynomial decay with $O(n^{-1})$.
\end{proof}

\begin{lem} \label{lem:GN1}
  Assume (without loss of generality) that $\E[X_j^4] \leq 4$ for all $j \in [p]$. For any $t > 0$, we have
  \begin{align}
    \Prob \left[ G_1(N) \geq 4 N k_n n^{-1/2} (1 + t)\right] \leq \exp \left( -2t^2/ K_n^2\right).
    \nonumber
  \end{align}
\end{lem}
\begin{proof}
  Recall that
  \begin{align}
    G_1(N) := \sup_{\gamma_j \in \B_j(N)} \left|(\P_n - \E) \left[ \ell \left( \X^T \hat{\bbeta} + Z_j \gamma^M_j, Y \right) - \ell \left( \X^T \hat{\bbeta} + Z_j \gamma_j, Y \right) \right] \indi_{n,j} \right|.  
    \nonumber
  \end{align}
  The proof follows the derivation of Lemma 5 in \citet{fan2010sure}.   
  First apply Lemma 1 in \citet{fan2010sure} \citep[Symmetrization, Lemma 2.3.1 in ][]{vaart1996weak} we obtain that
  \[
    \E G_1(N) \leq 2 \E \sup_{\gamma_j \in \B_j(N)} \left | \P_n\left \{
    \varepsilon \left[ \ell \left( \X^T \hat{\bbeta} + Z_j \gamma^M_j, Y \right) - \ell \left( \X^T \hat{\bbeta} + Z_j \gamma_j, Y \right) \right] \indi \left\{ (Z_j, Y) \in \Lambda_j\right\} \right\} \right|,
  \]
  where $\varepsilon_1, \ldots, \varepsilon_n$ is a Rademacher sequence independent of everything else. By Lemma 3 in \citet{fan2010sure} \citep[Contraction theorem in ][]{ledoux1991probability} and Condition \ref{condition:lip}, we further have
  \begin{align}
    \E G_1(N) \leq & 2 \E \sup_{\gamma_j \in \B_j(N)} \left| \P_n\left \{ \varepsilon k_n Z_j (\gamma_j - \gamma^M_j) \indi \left\{ (Z_j, Y) \in \Lambda_j\right\} \right\} \right| \nonumber\\
    \leq & 2 k_n \E \left| \P_n\left \{ \varepsilon Z_j  \indi \left\{ (Z_j, Y) \in \Lambda_j\right\} \right\} \right| \sup_{\gamma_j \in \B_j(N)} |\gamma_j - \gamma^M_j|\nonumber\\
    \leq & 2 k_n N\E \left| \P_n\left \{ \varepsilon Z_j  \indi \left\{ (Z_j, Y) \in \Lambda_j\right\} \right\} \right|, \nonumber
  \end{align}
  where
  \[
    \left(\E \left| \P_n\left \{ \varepsilon Z_j  \indi \left\{ (Z_j, Y) \in \Lambda_j\right\} \right\} \right|^2 \right)^{1/2} 
    = \left( n^{-2}\sum_{i = 1}^n \E[z_{ij}^2] * \indi \{ |z_{ij}| \leq K_n \}\right)^{1/2} \leq \sqrt{\E[Z_j^2]} n^{-1/2}.
  \]
  Assuming $\E[X_j^4] \leq 4$ for all $j \in [p]$, we have that $\sqrt{\E[Z_j^2]} \leq 2$. Thus using Jensen's inequality we have
  \begin{align}
    \E G_1(N) \leq 4 k_n N n^{-1/2}.
  \label{eq:concentration_expectation}
  \end{align}
  Next by Condition \ref{condition:lip} and Cauchy-Schwarz inequality,  on $\Lambda_j$, for any $\gamma_j \in \B_j(N)$, we have
  \begin{align}
    \left|\ell \left( \X^T \hat{\bbeta} + Z_j \gamma^M_j, Y \right) - \ell \left( \X^T \hat{\bbeta} + Z_j \gamma_j, Y \right) \right| \leq k_n | Z_j \left( \gamma_j - \gamma^M_j \right)| \leq k_n K_n N,
  \label{eq:concentration_bound}
  \end{align}
  Finally, given \eqref{eq:concentration_expectation} and \eqref{eq:concentration_bound}, we can infer from Lemma 4 in \citet{fan2010sure} \citep[Concentration theorem in][]{massart2000constants} that
  \begin{align}
    \Prob \left[ G_1(N) \geq 4k_n n^{-1/2} N (1 + t) \right] \leq \exp 
    \left(-2t^2/K^2 \right).
    \nonumber
  \end{align}
\end{proof}

\begin{lem} \label{lem:indi}
  For any $j \in [q]$, let \begin{align}
    \indi_{n, j} := \prod_{i = 1}^n \indi\left\{ (z_{ij}, y_i) \in \Lambda_j \right\},
    \nonumber
  \end{align}
  where $\Lambda_j$ is defined in \eqref{eq:lambdaj}.
  Under Condition \ref{condition:tail}, we have 
  \begin{align}
    \Prob \left( \indi_{n, j} = 0 \right) \leq n \left[ s_1 \exp (-r_0 K_n^\alpha) + 2 r_1 \exp (-r_0 K_n^{\alpha / 2})\right].
    \nonumber
  \end{align}

\end{lem}
\begin{proof}
  By definition in \eqref{eq:indi} and a union bound, we have
  \begin{align}
    \Prob \left( \indi_{n, j} = 0\right) 
    \leq \sum_{i = 1}^n \Prob \left( (z_{ij}, y_i) \notin \Lambda_j\right) 
    = n \Prob \left( (Z_j, Y) \notin \Lambda_j\right),
    \nonumber
  \end{align}
  where the equality holds since $(z_{ij}, y_i)$ are identically distributed for $i \in [n]$ and for any generic random vector $(Z_j, Y)$ following the same distribution. By the definition of $\Lambda_j$ in \eqref{eq:lambdaj}, we further have
  \begin{align}
    \Prob \left( (Z_j, Y) \notin \Lambda_j\right) \leq 
    \Prob \left( |Z_j| > K_n \right) + \Prob \left( |Y| > K^\ast_n \right).
    \nonumber
  \end{align}
  Note that for any interaction $Z_j = X_kX_\ell$.
  Condition \ref{condition:tail} implies that
  \begin{align}
    \Prob \left( |Z_j| > K_n \right) \leq \Prob \left(|X_k| > \sqrt{K_n}\right) + \Prob \left(|X_\ell| > \sqrt{K_n}\right) = 2r_1 \exp(-r_0 K_n^{\alpha / 2}).
    \nonumber
  \end{align}
  Finally, from Lemma \ref{lem:taily} and the definition that $K_n^\ast = r_0 K_n^\alpha / s_0$, we have
  \begin{align}
    \Prob \left(|Y| > K_n^\ast \right) \leq s_1 \exp (-r_0 K_n^\alpha).
    \nonumber
  \end{align}
\end{proof}

\section{Scale of $k_n$ and $K_n$}
\label{app:k_n and K_n}
\subsection{Choice of $k_n$}
\label{app:kn}
Recall that $\ell(\theta, y) = b(\theta) - \theta y$. For any $a_0 \in \B$, by the mean-value theorem,
\begin{align}
    &\left|\ell(a_0 + z_j \gamma_j, y) - \ell(a_0 + z_j \tilde{\gamma}_j, y) \right| \indi \left\{ (z_j, y) \in \Lambda_j \right\} \nonumber \\
  \leq & \max_{\check{\gamma}_j \in \B}\left(\left|b'(a_0+ z_j \check{\gamma}_j) \right| + |y| \right) \left| z_j \left( \gamma_j - \tilde{\gamma}_j \right) \right| \indi \left\{ (z_j, y) \in \Lambda_j \right\} \nonumber \\
  \leq & k_n \left| z_j \left( \gamma_j - \tilde{\gamma}_j \right) \right| \indi \left\{ (z_j, y) \in \Lambda_j \right\}, \nonumber
  \nonumber
\end{align}
where $k_n$ in \eqref{eq:kn} since $b'(\cdot)$ is an increasing function by Condition \ref{condition:lip} and the definition of $\B$.

\subsection{The order of dimensionality}
\label{app:Kn}
We adopt similar techniques in \citet{fan2010sure} and \citet{barut2016conditional} to derive the optimal scale of $K_n$.
Specifically, we need to balance the first and third terms in the probability upper bound in statement \texttt{S1} to minimize the upper bound. We then derive the dimensionality that the method can handle by ensuring the probability upper bound in statement \texttt{S1} converges to zero as $n$ approaches infinity.

In logistic regression models, $b^\prime(\cdot)$ is bounded, and thus the Lipschitz constant $k_n < 1/4 + r_0 K_n^\alpha / s_0$ can be regarded as a constant. Minimizing the probability upper bound by balancing the first and third terms gives the optimal order of $K_n = O(n^\frac{\min\{1 - 2\kappa, 2\rho - 2 \kappa\}}{2+\alpha/2})$. We then derive the dimensionality as $ \log(q) = o(n^\frac{\min\{1 - 2\kappa, 2\rho - 2 \kappa\}\alpha}{4+\alpha})$.
   
In linear models where $b^\prime(\theta) = \theta$, the Lipschitz constant $k_n = B (K_n + 1) + r_0K_n^\alpha / s_0$. Balancing the first and third terms in the probability upper bound and then substituting the value of $k_n$ gives:
\begin{align}
    O(K_n^{\alpha / 2}) &=     O(\frac{1}{K_n^2}\min \left\{ k_n^{-2}n^{1 - 2 \kappa}, n^{2 \rho - 2 \kappa} \right\}) \nonumber \\
    &= O\left(\frac{1}{K_n^2}\min \left\{ \frac{n^{1 - 2 \kappa}}{K_n^{\max\{2,2\alpha\}}}, n^{2 \rho - 2 \kappa} \right\}\right). \nonumber
\end{align}
Therefore, different choices of $\alpha$, $\kappa$, and $\rho$ can result in different optimal orders of $K_n$. This, in turn, affects the dimensionality that the method can handle:
for $\alpha \geq 1$,
\begin{align}
\log (q)= \begin{cases}
o(n^\frac{(1-2\kappa)\alpha}{4+5\alpha}) & \text{if }  \kappa \leq (\frac{1}{\alpha} + \frac{5}{4}) (\rho - \frac{1}{2}) + \frac{1}{2}\\
o(n^{\frac{(2\rho-2\kappa)\alpha}{4+\alpha}}) & \text{if } \kappa > (\frac{1}{\alpha} + \frac{5}{4}) (\rho - \frac{1}{2}) + \frac{1}{2}
\end{cases},
\nonumber
\end{align} 
and for $\alpha < 1$,
\begin{align}
\log (q)= \begin{cases}
o(n^\frac{(1-2\kappa)\alpha}{8+\alpha}) & \text{if } \kappa \leq (\frac{\alpha}{4}+2)(\rho - \frac{1}{2}) + \frac{1}{2}\\
o(n^{\frac{(2\rho-2\kappa)\alpha}{4+\alpha}}) & \text{if } \kappa > (\frac{\alpha}{4}+2)(\rho - \frac{1}{2}) + \frac{1}{2}\\
\end{cases}.
\nonumber
\end{align}

\section{Technical details in Section \ref{sec:linear_app}}\label{app:linear_app_proof}
For \eqref{eq:lin_baseline}, we have
\begin{align}
  \bbeta^M = \argmin_{\bbeta \in \real^p} \E \left[ (Y - \X^T \bbeta)^2 \right] 
  &= \Cov(\X)^{-1} \Cov(\X, \X^T \bbeta^\ast + \Z^T \bgamma^\ast + \varepsilon) \nonumber\\
  &= \bbeta^\ast + \bSigma^{-1} \bPhi \bgamma^\ast,
  \label{eq:lin_batahat_proof}
\end{align}
and for \eqref{eq:lin_population}
\begin{align}
  \gamma^M_j &= \argmin_{\gamma_j \in \real} \E \left[ \left( Y - \X^T \bbeta^M - Z_j \gamma_j \right)^2 \right] \nonumber\\
                 &= \argmin_{\gamma_j \in \real} \E \left[ \left( \left( \Z - \bPhi^T \bSigma^{-1} \X \right)^{T} \bgamma^\ast + \varepsilon - Z_j \gamma_j \right)^2 \right]  \nonumber\\
                 &= \Var(Z_j)^{-1} \Cov \left( Z_j, \left( \Z - \bPhi^T \bSigma^{-1} \X \right)^{T} \bgamma^\ast \right) \nonumber\\
                 &= \Psi_{jj}^{-1} \left[ \bPsi_{j \cdot} - \bPhi_{j \cdot} \bSigma^{-1} \bPhi \right] \bgamma^\ast. 
   %\label{eq:lin_gammabar_proof}
   \nonumber
\end{align}

For $\check{\gamma}_j^T$, we can derive
\begin{align}
  (\check{\bbeta}^T, \check{\gamma}_j)^T
  &\in \argmin_{\bbeta \in \real^p, \gamma_j \in \real} \E \left[ (Y - \X^T \bbeta - Z_j \gamma_j)^2 \right] \nonumber\\
  &= \left[\Cov \begin{pmatrix}
    \X \\
    Z_j
    \end{pmatrix} \right]^{-1} \Cov \left[ \begin{pmatrix}
    \X \\
    Z_j
\end{pmatrix}, \X^T \bbeta^\ast + \Z^T \bgamma^\ast + \varepsilon \right]  \nonumber \\
  &= \begin{pmatrix}
    \bSigma & \bPhi_{\cdot j} \\
    \bPhi_{j \cdot} & \Psi_{jj}
  \end{pmatrix}^{-1} \begin{pmatrix}
    \bSigma \bbeta^\ast + \bPhi \bgamma^\ast \\
    \bPhi_{\cdot j} \bbeta^\ast + \bPsi_{j \cdot} \bgamma^\ast
  \end{pmatrix}. 
  \nonumber
\end{align}
By the block matrix inversion formula, we get
\begin{align}
  \check{\gamma}_j &= - \left( \Psi_{jj} - \bPhi_{j \cdot} \bSigma^{-1} \bPhi_{\cdot j} \right)^{-1} \bPhi_{j \cdot} \bSigma^{-1} \left( \bSigma \bbeta^\ast + \bPhi \bgamma^\ast \right) + \left( \Psi_{jj} - \bPhi_{j \cdot} \bSigma^{-1} \bPhi_{\cdot j} \right)^{-1} \left( \bPhi_{\cdot j} \bbeta^\ast + \bPsi_{j \cdot} \bgamma^\ast \right) \nonumber\\
                   &= \left( \Psi_{jj} - \bPhi_{j \cdot} \bSigma^{-1} \bPhi_{\cdot j} \right)^{-1} \left( \bPsi_{j \cdot}  - \bPhi_{j \cdot} \bSigma^{-1} \bPhi \right)\bgamma^\ast. \label{eq:lin_gammacheck_proof}
\end{align}

Next, from \eqref{eq:lincov}, \eqref{def:cle_Y} and \eqref{eq:lin_batahat_proof}, we have
\begin{align}
  \Cov \left( Y, Z_j | \X \right) &= \E Z_j \left[ Y - \E_L \left( Y | \X \right) \right] = \E [Z_j Y] - \E \left[ Z_j \X^T \bbeta^M \right] \nonumber\\
                                  &= \E Z_j \left[ \X^T \bbeta^\ast + \Z^T \bgamma^\ast \right] - \E \left[ Z_j \X^T \left( \bbeta^\ast + \bSigma^{-1} \bPhi \bgamma^\ast \right) \right] \nonumber\\
                                  &= \left[ \bPsi_{j \cdot} - \bPhi_{j \cdot} \bSigma^{-1} \bPhi \right] \bgamma^\ast = \Psi_{jj} \gamma^M_j. 
                     \label{eq:lin_cov_proof}
\end{align}

\section{Screening property of $\hat{\mathcal{I}}_m^{\text {top }}$}\label{app:topm}
In this section, we will provide the theoretical guarantees of ``top-$m$'' strategy in Step 2 of \textit{sprinter} algorithm \ref{alg:sprinter}.
\begin{thm} \label{thm:toph}
Let 
\begin{align}
    \I_h^{\text{top}} =\left\{j \in[q]: \left|\gamma_j^{M} \right| \text { is among the top } h\text { largest}\right\}
    \label{toph}.
\end{align}
Under the conditions for Theorem \ref{thm:screening}, given that
\begin{align}
    \min _{j\in \mathcal{I}_h^{\text {top }}} \left|\gamma_j^{M} \right| \geq \max _{j \notin \mathcal{I}_h^{\text {top }}} \left|\gamma_j^{M} \right| + 2c_1 n^{-\kappa}
    \label{jump},
\end{align}
then for $m \geq h$, we have
\begin{align}
    \Prob \left(  \mathcal{I}_h^{\text {top }} \subseteq \hat{\mathcal{I}}_m^{\text {top }}   \right) 
    \geq &1 - |\A(\kappa)| \exp \left( - \frac{c_2}{K_n^2} \min \left\{ k_n^{-2} n^{1 - 2 \kappa}, n^{2 \rho - 2 \kappa} \right\}\right) \nonumber \\ & -  n|\A(\kappa)| \left[s_1 \exp (-r_0 K_n^\alpha) + 2 r_1 \exp (-r_0 K_n^{\alpha / 2})\right].
    \nonumber
\end{align}
\end{thm}
\begin{proof}

Recall $$
    \hat{\mathcal{I}}_m^{\text {top }} =\left\{j \in[q]: \left|\hat{\gamma}_j \right| \text { is among the top } m\text { largest}\right\}
$$
Here, $m \geq h$.

Let $\ell$ be 
$$\ell=\underset{j \in \hat{\mathcal{I}}_m^\text {top} \backslash \mathcal{I}_h^\text {top}}{\arg \min }\left|\hat{\gamma}_j\right|.$$
Here, $\ell$ denotes the index of the smallest magnitude of $\hat{\gamma}_j$ such that $\ell$ is included in $\hat{\mathcal{I}}_m^{\text {top }}$ but not included in $\mathcal{I}_h^{\text {top }}$.

If such $\ell$ does not exist, then it must hold that $m=h$ and $\hat{\mathcal{I}}_m^{\text {top }}=\mathcal{I}_h^{\text {top }}$, and the theorem holds. 

If such $\ell$ exists, then
\begin{align}
\min _{j \in \hat{\mathcal{I}}_m^{\text {top }}}\left|\hat{\gamma}_j\right| \leq\left|\hat{\gamma}_\ell\right| \leq\left|\gamma_\ell^M\right|+\left|\hat{\gamma}_\ell-\gamma_\ell^M\right| \leq \max _{j \notin \mathcal{I}_h^{\text {top }}}\left|\gamma_j^M\right|+\left|\hat{\gamma}_\ell-\gamma_\ell^M\right|.
\label{trineqr}
\end{align}

For any $j \in \mathcal{I}^{\text {top }}_h$, by triangle inequality,
\begin{align}
\left|\hat{\gamma}_j\right| \geq\left|\gamma_j^M\right|-\left|\hat{\gamma}_j-\gamma_j^M\right|.
\label{trineql}
\end{align}

Then, by subtracting \eqref{trineql} from \eqref{trineqr}, we get
\begin{align}
\left|\hat{\gamma}_j\right|-\min _{j \in \hat{\mathcal{I}}_m^{\text {top }}}\left|\hat{\gamma}_j\right| & \geq\left|\gamma_j^M\right|-\max _{j \notin \mathcal{I}_h^{\mathrm{top}}}\left|\gamma_\ell^M\right|-\left|\hat{\gamma}_j-\gamma_j^M\right|-\left|\hat{\gamma}_\ell-\gamma_\ell^M\right| \nonumber \\
&\geq \min _{j \in \mathcal{I}_h^{\text {top }}}\left|\gamma_j^M\right|-\max _{j \notin \mathcal{I}_h^{\mathrm{top}}}\left|\gamma_j^M\right|-2 \max _{j \in[q]}\left|\hat{\gamma}_j-\gamma_j^M\right| \nonumber \\
& \geq 0,
\nonumber
\end{align}
where the last inequality is derived from \eqref{jump} and Theorem \ref{thm:screening} with a certain probability. Hence, for any $j \in \mathcal{I}^{\text {top }}_h$, we can state that $j \in \hat{\mathcal{I}}_m^{\text {top }}$ holds true with the same probability as in Theorem \ref{thm:screening}. Thus, the theorem is proved.

\end{proof}

\section{A vectorization trick for OrdinalNet package}
\label{app:ordinet}
OrdinalNet \citep{wurm2017regularized} is a powerful package that use a coordinate descent algorithm to fit a wide class of ordinal regression models with an elastic-net penalty. In this section, we illustrate how to accelerate the computation process when applying it to a large-scale dataset. Those techinques are particularly applicable for the parallel form of ordinal regression, where all response categories share the same coefficients for predictors.

Suppose we have $n$ observations. The parallel form of the ordinal logistic regression model, with $k+1$ ordinal categories on the $i$-th observation, is defined as follows:
\begin{align}
    \operatorname{logit}(\bdelta_i) = \mathbf{X}_i \bbeta,
    \nonumber
\end{align}
Here, $\bdelta_i = \left(\delta_{i1}, \ldots, \delta_{ik}\right)^{T} \text{ where } \delta_{it} = P(Y_i \leq t) \text{ for all } t\in\{1,2,\cdots,k\}$. Additionally, $\mathbf{X}_i$ and $\bbeta$ is defined as:
\begin{align}
\mathbf{X}_i  =\left(\begin{array}{l|l}
\bm{I_k} & \begin{array}{c}
\bm{x}_i^{T} \\
\vdots \\
\bm{x}_i^{T}
\end{array}
\end{array}\right)_{k \times(k+p)}, \quad \bbeta=\left(\begin{array}{l}
\bm{c} \\
\bm{b}
\end{array}\right)_{(k+p) \times 1},
\nonumber
\end{align}
and $\bm{x}_i\in \real^p$ is a $p$-dimensional feature vector without intercept. Therefore, different categories share the same coefficient vector $\bm{b}\in \real^p$, but each has a unique intercept term $c_t$ in $\bm{c} = \{c_1,c_2, \cdots, c_k\}^{T}\in \real^k$, with $c_t$ being the intercept of the $t$-th category. 

The elastic-net penalized objective function is defined as
\begin{align}
\mathcal{M}\left(\bbeta ; \alpha, \lambda, \omega_1, \ldots, \omega_{k+p}\right)=-\frac{1}{n} \ell(\bbeta)+
\lambda\left\{\alpha\|\bm{b}\|_1+(1-\alpha)\|\bm{b}\|_2^2\right\},
\nonumber
\end{align}
where the first term is the log-likelihood and the second term is the elastic-net penalty applied to the coefficient vector $\bm{b}$. Here, $\lambda$ is a regularization parameter, and $\alpha$ (varying from 0 to 1) determines the relative weight of the $\ell_1$-penalty and $\ell_2$-penalty.

Let $\boldsymbol{p}_i$ be the ordinal category probability vector for the $i$-th observation, and use $h$ to denote the logit link inverse function from $\boldsymbol{\eta}_i=\mathbf{X}_i\bbeta$ to $\boldsymbol{p}_i$. Let $L_i(\boldsymbol{p}_i)$ represent the log-likelihood of the $i$-th observation with the probability vector $\boldsymbol{p}_i$, then we have
\begin{align}
\ell_i(\bbeta)=L_i(\boldsymbol{p}_i)=L_i\left(h\left(\boldsymbol{\eta}_i\right)\right)= L_i\left(h\left(\mathbf{X}_i\bbeta\right)\right).
\nonumber
\end{align}

The optimization applies a Taylor expansion to update the quadratic approximation to $\ell(\bbeta)$. This requires the score function and Fisher information for $\ell_i(\bbeta)$.

By a chain rule decomposition, we have score function
\begin{align}
U_i(\bbeta)=\nabla \ell_i(\bbeta)=\mathbf{X}_i^{T} D h\left(\boldsymbol{\eta}_i\right)^{T} \nabla L_i\left(\boldsymbol{p}_i\right),
\label{score}
\end{align}
where $\nabla$ is the gradient operator and $D$ is the Jacobi operator.

The Fisher information matrix is
\begin{align}
\mathcal{I}_i(\bbeta)=E_{\bbeta}\left(U_i(\bbeta) U_i(\bbeta)^{\top}\right)=\mathbf{X}_i^{T} \W_i \mathbf{X}_i,
\label{info}
\end{align}
where
\begin{align}
\W_i=D h\left(\boldsymbol{\eta}_i\right)^{T} E_{\bbeta}\left\{\nabla L_i\left(\boldsymbol{p}_i\right) \nabla L_i\left(\boldsymbol{p}_i\right)^{T}\right\} D h\left(\boldsymbol{\eta}_i\right). 
\nonumber
\end{align}

Because of the independence of observations, the full data score and information would be
\begin{align}
U(\bbeta)=\sum_{i=1}^n U_i(\bbeta) \quad \text {and} \quad \mathcal{I}(\bbeta)=\sum_{i=1}^n \mathcal{I}_i(\bbeta).
\label{summ}
\end{align}
As in \eqref{summ}, the ordinalNet package uses a for loop in R to pass through all data to evaluate the score and information for each observation and summing them up. However, for datasets with a large number of observations, like the Tripadvisor dataset in Section \ref{sec:Tripadvisor}, using a for loop in R becomes computationally inefficient. Instead of using a for loop, we vectorize the score and information and employ block-diagonal sparse matrix multiplication, which significantly reduces the computational burden.

Here, we rewrite \eqref{summ} as
\begin{align}
U(\bbeta)=\mathbf{X}^{T} [\mathbf{Q}^{T} \mathbf{P}] = \mathbf{X}^{T}\mathbf{V},
%\label{score matrix}
\nonumber
\end{align}
where 
\begin{align}
\nonumber
\mathbf{X}=\left(\begin{array}{c}
\mathbf{X}_1 \\
\vdots \\
\mathbf{X}_n
\end{array}\right)_{nk \times (k+p)} , 
\quad
\mathbf{Q}=\left(\begin{array}{cccc}
D h\left(\boldsymbol{\eta}_1\right) & 0 & \cdots & 0 \\
0 & D h\left(\boldsymbol{\eta}_2\right) & \cdots & 0 \\
\vdots & \vdots & \ddots & \vdots \\
0 & 0 & \cdots & D h\left(\boldsymbol{\eta}_n\right)
\end{array}\right)_{nk \times nk}, 
\nonumber
\\
\mathbf{P}=\left(\begin{array}{c}
\nabla L_1\left(\boldsymbol{p}_1\right) \\
\vdots \\
\nabla L_n\left(\boldsymbol{p}_n\right) 
\end{array}\right)_{nk \times 1},
\quad 
\mathbf{V}=\left(\begin{array}{c}
\boldsymbol{V}_1 \\
\vdots \\
\boldsymbol{V}_n
\end{array}\right)
= \left(\begin{array}{c}
D h\left(\boldsymbol{\eta}_1\right)^{T} \nabla L_1\left(\boldsymbol{p}_1\right) \\
\vdots \\
D h\left(\boldsymbol{\eta}_n\right)^{T} \nabla L_n\left(\boldsymbol{p}_n\right)
\end{array}\right)_{nk \times 1},
\nonumber
\end{align}
and
\begin{align}
\mathcal{I}(\bbeta)=\mathbf{X}^{T} \mathbf{W} \mathbf{X},
%\label{info matrix}
\nonumber
\end{align}
where
\begin{align}
\mathbf{W}=\left(\begin{array}{cccc}
\boldsymbol{W}_1 & 0 & \cdots & 0 \\
0 & \boldsymbol{W}_2 & \cdots & 0 \\
\vdots & \vdots & \ddots & \vdots \\
0 & 0 & \cdots & \boldsymbol{W}_n
\end{array}\right)_{nk \times nk}.
\nonumber
\end{align}

To reduce memory usage, we can use ``dgCMatrix" in R to represent sparse matrices $\mathbf{Q}$ and $\mathbf{W}$. Then the computation efficiency can then be greatly reduced by sparse matrix multiplication.

We can further improve computation efficiency by using the special structure of the $\mathbf{X}_i$ matrix. Recall that, 
\begin{align}
\mathbf{X}_i=\left(\begin{array}{l|l}
\bm{I}_{k} & \begin{array}{c}
\bm{x}_i^{T}  \\
\vdots \\
\bm{x}_i^{T} 
\end{array}
\end{array}\right)
= \Bigl(\bm{I}_{k}, \boldsymbol{1}_k \bm{x}_i^{T} \Bigr),
\nonumber
\end{align}
then \eqref{score} and \eqref{info} will become 
\begin{align}
U_i(\bbeta)=\mathbf{X}_i^{T} D h\left(\boldsymbol{\eta}_i\right)^{T} \nabla L_i\left(\boldsymbol{p}_i\right)
=\left(
\begin{array}{c}
\bm{I_k} \\
\bm{x}_i\boldsymbol{1}_k^{T}
\end{array}\right)
\boldsymbol{V}_i = 
\left(
\begin{array}{c}
\boldsymbol{V}_i \\
\bm{x}_i^{T}[\boldsymbol{1}_k^{T}\boldsymbol{V}_i]
\end{array}\right),
\nonumber
\end{align}
and
\begin{align}
\mathcal{I}_i(\bbeta)
=\mathbf{X}_i^{T} \W_i \mathbf{X}_i 
= \left(\begin{array}{cc}
\W_i & [\W_i \boldsymbol{1}_k] \bm{x}_i^{T} \\
\bm{x}_i [\W_i \boldsymbol{1}_k]^{T} & \bm{x}_i[\boldsymbol{1}_k^{T}\W_i\boldsymbol{1}_k] \bm{x}_i^{T}
\end{array}\right).
\nonumber
\label{info}
\end{align}
This can significantly lessen the computational burden. In the Tripadvisor experiment with training set $n = 21132$ and $p = 7817$, the time to obtain the score and information is significantly reduced from 2 hours to 5 seconds (on a 2020 Macbook Pro. 13.3 inch, Apple M1 chip, RAM 8GB), which is quite impressive. 

\section{Additional results for the Tripadvisor dataset application}
\label{app:tripresult}
In this section, we show the top 10 main effects estimation and top 10 interactions estimation from the application of the \texttt{sprinter} algorithm to the Tripadvisor hotel reviews dataset.
\begin{table}[h!]
\begin{minipage}[b]{0.45\linewidth}
\centering
\begin{tabular}{cc}
    \hline \text { word } & $ \text{coefficient} $ \\
    \hline \text { wonderful } & 1.413 \\
    \text { excellent } & 1.300 \\
    \text { fantastic } & 1.297 \\
    \text { loved } & 1.278 \\
    \text { perfect } & 1.110 \\
    \text { amazing } & 1.014 \\
    \text { magical } & 0.918 \\
    \text { outstanding } & 0.877 \\
    \text { superb } & 0.809 \\
    \text { best } & 0.797\\
\hline 
\end{tabular}
\end{minipage}
\hspace{0.5cm}
\begin{minipage}[b]{0.45\linewidth}
\centering
    \begin{tabular}{cc}
    \hline \text { word } & \text{coefficient} \\
    \hline  \text { worst } & -2.583 \\
    \text { terrible } & -1.958 \\
    \text { dirty } & -1.685 \\
    \text { not } & -1.328 \\
    \text { horrible } & -1.268 \\
    \text { charged } & -1.051 \\
    \text { poor } & -1.021 \\
    \text { rude } & -0.970 \\
    \text { awful } & -0.946 \\
    \text { disgusting } & -0.867\\
    \hline 
    \end{tabular}
\end{minipage}
\caption{Top 10 recovered positive main effects estimates (left) and top 10 recovered negative main effects estimates (right)}
\label{tab:main}
\end{table}

\begin{table}[h!]
\centering
\begin{tabular}{ccccc}
\hline \text{word1} &  \text{word2}& \text{word1$\times$word2 coefficient} & \text{word1 coefficient}  & \text{word2 coefficient} \\
\hline \text { great } & { fantastic } & -0.592 & 0.694 & 1.297 \\
\text { fabulous} & {perfect } & -0.513 & 0.677 & 1.110 \\
\text { free } & { positive } & -0.502 & 0.244 & -0.118 \\
\text { great } & {loved } & -0.494 & 0.694 & 1.278 \\
\text { outstanding } & { beautiful } & -0.451 & 0.877 & 0.199\\
\hline 
\end{tabular}
\caption{Top 5 negative interaction estimates and their corresponding main effects estimates}
\label{tab:posi inter}
\end{table}

\begin{table}[h!]
\centering
\begin{tabular}{cccccc}
\hline \text{word1} &  \text{word2}& \text{word1$\times$word2 coefficient} & \text{word1 coefficient}  & \text{word2 coefficient} \\
\hline \text { worst } & \text { but } & 0.618 & -2.583 & -0.577 \\
\text { good } & \text { charged } & 0.508 & -0.245 & -1.050 \\
\text { great } & \text { horrible } & 0.491 & 0.694 & -1.268 \\
\text { not } & \text { but } & 0.482 & -1.328 & -0.577\\
\text { good } & \text { terrible } & 0.459 & -0.245 & -1.958\\
\hline 
\end{tabular}
\caption{Top 5 positive interaction estimates and their corresponding main effects estimates}
\label{tab:nega inter}
\end{table}

\end{document}